\documentclass[12pt,letterpaper]{article}
\usepackage[utf8]{inputenc}
\usepackage[english]{babel}
\usepackage{amsmath}
\usepackage{amssymb}
\usepackage{amsfonts}
\usepackage{graphicx}
\usepackage{bm}
\usepackage{setspace}
\usepackage{sectsty}
\usepackage[position=top]{subfig}
\usepackage[round]{natbib}
\usepackage[margin=1.25in]{geometry}
\usepackage{pict2e}

\sectionfont{\nohang\centering\mdseries\sffamily}
\subsectionfont{\sffamily\mdseries}
\title{Random Walks on Directed Networks: Inference and Respondent-driven Sampling}

\author{Jens Malmros\\ Department of Mathematics, Stockholm University,\\ Stockholm, Sweden \and Naoki Masuda\\Department of Mathematical Informatics, University of Tokyo,\\Tokyo, Japan \and Tom Britton\\ Department of Mathematics, Stockholm University,\\ Stockholm, Sweden\footnote{Jens Malmros is Ph.D. student, Department of Mathematics, Division of Mathematical Statistics, Stockholm University, SE-106 91 Stockholm (E-mail: jensm@math.su.se). Naoki Masuda is associate professor, Department of Mathematical Informatics, The University of Tokyo, 7-3-1 Hongo, Bunkyo, Tokyo 113-8656, Japan (E-mail: masuda@mist.i.u-tokyo.ac.jp). Tom Britton is professor, Department of Mathematics, Division of Mathematical Statistics, Stockholm University, SE-106 91 Stockholm (E-mail: tomb@math.su.se). J.M. was supported by grant no.~2009-5759 from the Swedish Research Council. N.M. was supported by Grants-in-Aid for Scientific Research (No.~23681033) from MEXT, Japan, the Nakajima Foundation, and the Aihara
Project, the
FIRST program from JSPS, initiated by CSTP, Japan.
The authors would like to thank prof. Fredrik Liljeros and dr. Xin Lu, Department of Sociology, Stockholm University for use of the Qruiser dataset.}}

\begin{document}
\maketitle
\newpage
\abstract{Respondent driven sampling (RDS) is a method often used to estimate
population properties (e.g.\ sexual risk behavior) in hard-to-reach populations.
It combines an effective modified snowball sampling methodology with an
estimation procedure that yields unbiased population estimates under the
assumption that the sampling process behaves like a random walk on the social
network of the population. Current RDS estimation methodology assumes that the
social network is undirected, i.e.\ that all edges are reciprocal. However,
empirical social networks in general also have non-reciprocated edges. To
account for this fact, we develop a new estimation method for RDS in the
presence of directed edges on the basis of random walks on directed networks. We
distinguish directed and undirected edges and consider the possibility that the
random walk returns to its current position in two steps through an undirected
edge. We derive estimators of the selection probabilities of individuals as a
function of the number of
outgoing edges of sampled individuals. We evaluate the performance of the proposed estimators on artificial and
empirical networks to show that they generally perform better than existing methods. This is in particular the case when
the fraction of directed edges in the network is large.} 

\newpage
\section{INTRODUCTION}\label{Sec:Introduction}

Random walks on networks are crucial to the understanding of many network processes, and in many applications, random walks serve as either rigorous or approximate tools depending on the amount of information available about networks. A network sampling methodology taking advantage of a random walk approximation is respondent-driven sampling (RDS). The method, first suggested in \citet{Heckathorn1997}, is especially suitable for investigating hidden or hard-to-reach populations, such as injecting drug users (IDUs), sex workers, and men who have sex with men (MSM). For such populations, sampling frames are typically unavailable because individuals often suffer from social stigmatization and/or legal difficulties, and conventional sampling methods therefore fail. High demand for valid inference on hidden populations, e.g.\ on the risk behavior of individuals and the disease prevalence in the population, as well as a lack of competing methods, has made RDS a leading method. Examples of RDS studies from 2013
include MSM in Nanjing, China~\citep{Tang2013}, undocumented Central American immigrants in Houston, Texas~\citep{Montealegre2013}, and IDUs in the District of Columbia~\citep{Magnus2013}.

At the core of RDS is the notion of a social network that binds the population together. During the sampling process, already sampled individuals use their social relations (edges of the social network) to recruit new individuals in the population into the sample, creating a snowball-like mechanism. Additionally, information on the structure of the network collected during the sampling process facilitates unbiased population estimates given that the actual RDS recruitment process behaves like a random walk on the network \citep{Salganik2004,Volz2008}.

In recent years, much RDS research has focused on the sensitivity of current RDS
estimators to violations of the assumptions underlying the estimating process.
In fact, it has been shown that RDS estimators may be subject to substantial
biases and large variances when some assumptions are not valid
\citep{Gile2010SocMet,Lu2012JRSS,Wejnert2009,Tomas2011EJS,Goel2010}. New RDS
estimators have been developed to mitigate this problem
\citep{Gile2011arXiv,Gile2011JASA,LuEtal}.

Current RDS estimation assumes that the social network of the population is undirected. However, real social networks are at least partially directed in general. The directedness of a network can be quantified by the the ratio of the number of non-reciprocal (i.e., directed) edges to the total number of edges in the network \citep{Wasserman1994book}. This value lies between 0 and 1, and a large value indicates that the network is close to a purely directed network. Examples of real social networks and social networks, including e-mail social networks, from online communities having a considerable fraction of non-reciprocal edges are shown in Table \ref{Tab:SocNetworks}. For these and other directed social networks, RDS methods assuming an undirected network may be biased.

\begin{table}[h]
\setlength{\tabcolsep}{5pt}
\centering
\small
\caption{Proportion of directed edges in social networks.}\label{Tab:SocNetworks}
\begin{tabular}{@{}ll@{\hspace{1cm}}ll@{}}
\hline\hline
\textbf{Real social networks} & & \textbf{Online social networks} &\\
\hline
High-tech managers & 0.71 & Google+ (Oct 2011) & 0.62 \\
\citep{Wasserman1994book} & & \citep{GongEtal2013} &\\
Dining partners & 0.76 & Flickr (May 2007) & 0.55\\
\citep{Moreno} & & \citep{GongEtal2013}&\\
Radio amateurs & 0.59 & LiveJournal (Dec 2006) & 0.26\\
\citep{Killworth1976} & & \citep{Mislove} & \\
& & Twitter (June 2009) & 0.78\\
& & \citep{Kwak} &\\
& & University e-mail & 0.77\\
& & \citep{Newman2002} &\\
& & Enron e-mail & 0.85\\
& & \citep{Boldi2004}&\\
& & \citep{Boldi2011}&\\
\end{tabular}
\end{table}

Motivated by these data, we aim to expand RDS estimation to the case of directed networks. Because the RDS method uses the random walk, a random walk framework for directed networks is a key component to this expansion. This is not a trivial task because the random walk behaves very differently in undirected and directed networks. In particular, the stationary distribution of the random walk is simply proportional to the degree of the vertex in undirected networks \citep{Doyle1984book,Lovasz1993Boyal}, whereas it is affected by the entire network structure in directed networks \citep{Donato2004EPJB,Langville2006book,MasudaOhtsuki2009NewJPhys}.

In this paper, we first present the commonly available RDS estimation procedures and the basics of random walks on networks in Sections \ref{Sec:RDS} and \ref{Sec:RandomWalks}, respectively. Then, we present methods for estimating the stationary distribution from random walks on directed networks and its application to RDS estimation in Section \ref{Sec:Estimation}. These methods are then evaluated and compared to existing methods by numerical simulations, which we describe in Section~\ref{Sec:Simulation}. The results from simulations are presented in Section~\ref{Sec:Results}. Finally, our findings are discussed in Section~\ref{Sec:Discussion}.

\section{RESPONDENT-DRIVEN SAMPLING}\label{Sec:RDS}

In practice, an RDS study begins with the selection of a seed group of
individuals from the population. Each seed is given a fixed number of coupons,
typically three to five, which are effectively the tickets for participation in
the study, to be distributed to other peers in the population. Those who have
received a coupon and joined the study (i.e., respondents) are also given
coupons to be distributed to other peers that have not obtained a coupon. This
procedure is repeated until the desired sample size has been reached. Each
respondent is rewarded both for participating in the study and for the
participation of those to whom he/she passed coupons, resulting in double
incentives for participation. The sampling procedure ensures that the identities
of members of the population are not revealed in the recruitment process. For
each respondent, the properties of interest (e.g., HIV status), number of
neighbors (degree), and the neighbors that the respondent has successfully
recruited are recorded.

We approximate the RDS recruitment process by a random walk on the social network. To this end, we assume that (i) respondents recruit peers from their social contacts with uniform probability, (ii) each recruitment consists of only one peer, (iii) sampling is done with replacement, such that a respondent may appear in the sample multiple times, (iv) the degree of respondents is accurately reported, and (v) the population forms a connected network. Then, if the random walk is in equilibrium with a known stationary distribution $\{\pi_i; i=1,\dots,N\}$, where $N$ is the population size, we may estimate $p_A$, the fraction of individuals with a property of interest $A$, as~\citep{ThompsonSampling}
\begin{equation}\hat{p}_A = \frac{\sum_{i\in S\cap A}1/\pi_i}{\sum_{i\in S}1/\pi_i},\label{Eq:GenProbEst}\end{equation}
where $S$ is our sample. For undirected networks, the stationary distribution is proportional to the degree \citep{Doyle1984book,Lovasz1993Boyal}, and Eq.\ \eqref{Eq:GenProbEst} yields the most widely used RDS estimator \citep{Volz2008} given by
\begin{equation}\hat{p}_A^{\rm{VH}} = \frac{\sum_{i\in S\cap A}1/ d_i}{\sum_{i\in S}1/ d_i},\label{Eq:RDSEst}\end{equation}
where $d_i$ is the degree of node $i$. However, the estimator given by
Eq.~\eqref{Eq:RDSEst} may be biased for directed networks
\citep{Lu2012JRSS,LuEtal}. Therefore, to estimate $p_A$ without bias from an RDS
sample on a directed network, we need to accurately calculate
Eq.~\eqref{Eq:GenProbEst}. Because the stationary distribution $\{\pi_i\}$ used
in Eq.~\eqref{Eq:GenProbEst} is analytically intractable for most directed
networks, we will proceed by deriving estimators of it.

\section{RANDOM WALKS ON DIRECTED NETWORKS}\label{Sec:RandomWalks}

We consider a directed, unweighted, aperiodic, and strongly connected network $G$ with $N$ vertices. Let $e_{ij}=1$ if there is a directed edge from $i$ to $j$ and 0 otherwise. An undirected edge exists between $i$ and $j$ if and only if $e_{ij}=e_{ji}=1$. We denote the number of undirected, in-directed, and out-directed edges at vertex $i$ by $d_i^{\rm{(un)}}$, $d_i^{\rm{(in)}}$, and $d_i^{\rm{(out)}}$, respectively. We use
$D^{\rm{(un)}}$, $D^{\rm{(in)}}$, and $D^{\rm{(out)}}$ to refer to the corresponding random variables if a node is drawn uniformly at random.
If we specifically mention that the network is undirected, we obtain $d_i^{\rm{(in)}}=d_i^{\rm{(out)}}=0$, and the degree of vertex $i$ refers to $d_i^{\rm{(un)}}=d_i$. Otherwise, the degree of vertex $i$ refers to the triplet
$(d_i^{\rm{(un)}}, d_i^{\rm{(in)}}, d_i^{\rm{(out)}})$.
We refer to $d_i^{\rm{(un)}}+d_i^{\rm{(in)}}$ and
$d_i^{\rm{(un)}}+d_i^{\rm{(out)}}$ as the in-degree and out-degree of vertex
$i$, respectively. It should be noted that we may observe for example the
out-degree $d_i^{\rm{(un)}}+d_i^{\rm{(out)}}$, but not separately the
$d_i^{\rm{(un)}}$ and $d_i^{\rm{(out)}}$ values.

Consider the simple random walk $\bm X=\{X(t);t=0,1,\ldots\}$ with state space $\bm S=\{1,\dots,N\}$ on $G$ such that the walker staying at vertex $i$ moves to any of the $d_i^{\rm{(un)}}+d_i^{\rm{(out)}}$ neighbors reached by an undirected or out-directed edge with equal probability. We denote the stationary
distribution of $\bm X$ by $\{\pi_i;i=1,\ldots,N\}$, where $\pi_i=\lim_{t\to\infty} P(X(t)=i)$. If we sample from the random walk in equilibrium, vertices will be selected with probabilities given by the stationary distribution, and we then refer to $\{\pi_i\}$ as the \emph{selection probabilities} of the vertices in $G$.

For an arbitrary network, we obtain
\begin{equation}
\pi_i = \sum_{j=1}^N \frac{e_{ji}}{\sum_{\ell=1}^N e_{j\ell}}\pi_j
= \sum_{j=1}^N \frac{e_{ji}}{d_j^{\rm{(un)}}+d_j^{\rm{(out)}}}\pi_j,
\label{Eq:GenSampProbs}
\end{equation}
where the stationary distribution is fully defined by $\sum_{i=1}^N \pi_i=1$. In undirected networks, we obtain $\pi_i = d_i/\sum_{j=1}^N d_j$. In contrast, there is no analytical closed form solution for $\{\pi_i\}$ in directed networks. If a directed network has little assortativity (i.e., degree correlation between adjacent vertices), $\{\pi_i\}$ is often accurately estimated by the normalized in-degree \citep{LuEtal,FortunatoEtal,Ghoshal2011NatComm} because
\begin{equation}
\pi_i \approx \sum_{j=1}^N \frac{e_{ji}}{d_j^{\rm{(un)}}+d_j^{\rm{(out)}}}\bar{\pi}
\propto \sum_{j=1}^N e_{ji} = d_i^{\rm{(in)}}+d_i^{\rm{(un)}},\label{Eq:EstPropIndeg}
\end{equation}
where $\bar{\pi}$ is the average selection probability. However, the estimate
given by~\eqref{Eq:EstPropIndeg} is often inaccurate in general directed
networks \citep{Donato2004EPJB,MasudaOhtsuki2009NewJPhys}. Moreover, since it is
much easier for individuals to assess how many people they know (i.e.,
out-degree) than by how many people they are known (i.e., in-degree), it is
common to observe only the out-degree. In this case, Eq.~\eqref{Eq:EstPropIndeg}
can not be used with an RDS sample.

\section{ESTIMATION OF SELECTION PROBABILITIES FOR DIRECTED NETWORKS}\label{Sec:Estimation}

We now derive estimators of the selection probabilities for the random walk on directed networks. We first derive an estimation scheme when the full degree $(d_i^{\rm{(un)}}, d_i^{\rm{(in)}}, d_i^{\rm{(out)}})$ is observed for all the vertices $i$ visited by the random walk. Then, we extend this estimation to the situation in which only the out-degree $d_i^{\rm{un}}+d_i^{\rm{out}}$ of the visited vertices is observed.

\subsection{Estimating Selection Probabilities From Full Degrees}\label{SubSec:SelectionProbabilities}

In order to estimate $\{\pi_i\}$, we assume that $X(t_0)=i$ and that $t_0$ is sufficiently large for the stationary distribution to be reached. We evaluate the frequency with which $\bm X$ visits $i$ in the subsequent times. If $\bm X$ leaves $i$ through an undirected edge $e_{i\cdot}^{\rm{(un)}}$, where $e_{i\cdot}^{\rm{(un)}}$ is one of the $d_i^{\rm{(un)}}$ undirected edges owned by $i$, $\bm X$ may return to $i$ after two steps using the same edge and repeat the same type of returns $m$ times in total, perhaps using different undirected edges $e _{i\cdot}^{\rm{(un)}}$. Then, $X(t_0)=X(t_0+2)=\cdots = X(t_0+2m)=i$ and $X(t_0+2m+2)=k$ for some $k\neq i$.

If $X(t_0+2)=i$, the walk first moves from $i$ through an undirected edge to vertex $j$ at $t=t_0+1$ and returns to $i$ through the same edge at $t=t_0+2$. The probability of this event is given by $d_i^{\rm{(un)}}/(d_i^{\rm{(un)}}+d_i^{\rm{(out)}})\cdot 1/(d_j^{\rm{(un)}}+d_j^{\rm{(out)}})$. Because the out-degree of vertex $j$, i.e., $d_j^{\rm{(un)}}+d_j^{\rm{(out)}}$, is unknown, we approximate $1/(d_j^{\rm{(un)}}+d_j^{\rm{(out)}})$ by $E\left(1/(\tilde D^{\rm{(un)}}+D^{\rm{(out)}})\right)$. Here $\tilde D^{\rm{(un)}}$ denotes the undirected degree distribution under the condition that the vertex is reached by following an undirected edge, i.e.\ a \emph{size-biased} distribution for the undirected degree,
$P(\tilde D^{\rm{(un)}}=d)\propto dP(D^{\rm{(un)}}=d)$~\citep{Newman2010}. It
is also possible to estimate $1/(d_j^{\rm{(un)}}+d_j^{\rm{(out)}})$ by
$1/E(\tilde D^{\rm{(un)}}+D^{\rm{(out)}})$, which however showed to have
hardly any effect in our simulations, and if any, slightly worse. Thus, we
estimate
the probability of returning to vertex $i$ after two steps by
\begin{equation}
p_i^{\rm{(ret)}}=\frac{d_i^{\rm{(un)}}}{d_i^{\rm{(un)}}+d_i^{\rm{(out)}}}E\left(\frac{1}{\tilde D^{\rm{(un)}}+D^{\rm{(out)}}}\right).\label{Eq:p-k-ret}
\end{equation}

When $t\ge t_0+2m+3$, we use Eq.~\eqref{Eq:EstPropIndeg} to estimate the probability to visit vertex $i$ at any time as being proportional to $d_i^{\rm{(un)}}+d_i^{\rm{(in)}}$, i.e.,
\begin{equation}
p_i^{\rm{(vis)}}=\frac{d_i^{\rm{(un)}}+d_i^{\rm{(in)}}}{N(E(D^{\rm{(un)}})+E(D^{\rm{(in)}}))}.\label{Eq:p-k-vis}
\end{equation}

Under these estimates, the number of returns after two steps to vertex $i$,
counting the starting point $X(t_0)=i$ as a return to $i$, is geometrically
distributed with expected value $1/(1-p_i^{\rm{(ret)}})$, and the number of
steps starting from $t=t_0+2m+2$, counting this step, and ending at the time
immediately before visiting $i$ with probability $p_i^{\rm{(vis)}}$ is
geometrically distributed with expected value $1/p_i^{\rm{(vis)}}$.
\begin{figure}
\centering
\subfloat[$Z_i^n$ consecutive returns to $i$.]{
\setlength{\unitlength}{1cm}
\begin{picture}(8,4)(-1,-1)
\thicklines
% First graph
\put(1,1){\circle*{0.25}}
\put(-0.1,2.2){\circle*{0.25}}
\put(2.1,2.2){\circle*{0.25}}
\put(1,-.4){\circle*{0.25}}
\put(1.2,1.2){\vector(1,1){0.8}}
\put(2,2){\vector(-1,-1){0.8}}
\put(0,2){\vector(1,-1){0.8}}
\put(1,.8){\vector(0,-1){1}}
\put(0.95,1.25){$i$}
\put(2.05,2.5){$j$}

% Second graph
\put(5,1){\circle*{0.25}}
\put(3.9,2.2){\circle*{0.25}}
\put(6.1,2.2){\circle*{0.25}}
\put(5,-.4){\circle*{0.25}}
\put(5.2,1.2){\vector(1,1){0.8}}
\put(6,2){\vector(-1,-1){0.8}}
\put(4,2){\vector(1,-1){0.8}}
\put(5,.8){\vector(0,-1){1}}
\put(4.95,1.25){$i$}
\put(6.05,2.5){$j$}
\put(2.7,1){\vector(1,0){.8}}
\put(3.5,1.2){\vector(-1,0){.8}}
\thinlines
\put(1.8,1.4){\vector(1,1){0.4}}
\put(6.2,1.8){\vector(-1,-1){0.4}}
\end{picture}}\qquad
\centering
\subfloat[Leaves $i$ for $Y_i^n$ steps.]{
\setlength{\unitlength}{1cm}
\begin{picture}(5,4)(-1,-1)
\thicklines
\put(1,1){\circle*{0.25}}
\put(-0.1,2.2){\circle*{0.25}}
\put(2.1,2.2){\circle*{0.25}}
\put(1,-.4){\circle*{0.25}}
\put(1.2,1.2){\vector(1,1){0.8}}
\put(2,2){\vector(-1,-1){0.8}}
\put(0,2){\vector(1,-1){0.8}}
\put(1,.8){\vector(0,-1){1}}
\put(0.95,1.25){$i$}
\put(2.05,2.5){$j$}
\thinlines
\put(1.2,0.6){\vector(0,-1){.5}}
\multiput(2.3,2.4)(0.2,0.2){3}{\line(1,1){0.1}}
\multiput(1.9,2.4)(-0.2,0.2){3}{\line(-1,1){0.1}}
\put(2.7,2.5){\vector(1,1){.25}}
\put(1.5,2.5){\vector(-1,1){.25}}
\end{picture}}
\caption{Schematic of a renewal period. a) The walker makes $Z_i^n$ consecutive
direct returns to $i$. b) The walker leaves $i$ without an immediate return,
because the walker leaves $i$ by a directed edge or leaves $j$ by another edge.
Then, the walker returns to $i$ after $Y_i^n$ steps.}\label{Fig:Schematic}
\end{figure}
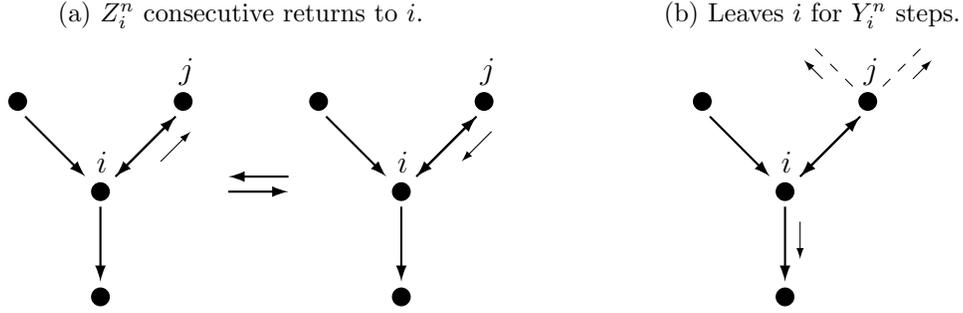
We then have a renewal process $\{R_i^n; n\ge 1, R_i^0=0\}$ with the $n$th renewal occurring at random time $R_i^n=\sum_{k=1}^n(2Z_i^{k}+Y_i^{k})$, where $Z_i^n\sim Ge(1-p_i^{\rm{(ret)}})$ and $Y_i^n\sim Ge(p_i^{\rm{(vis)}})$. In Figure~\ref{Fig:Schematic}, the behavior of the process during a renewal period is schematically shown. The average time step between consecutive renewal events is equal to $2E(Z_i^n)+E(Y_i^n)$. The average number of visits to $i$ between the two renewal events, with the visit to $i$ at $t=t_0$ included, is equal to $E(Z_i^n)$. Therefore, from renewal theory \citep[see e.g.,][]{Resnick}, we obtain an estimate of $\pi_i$ as
\begin{equation}
\pi_i\approx \frac{E(Z_i^n)}{2E(Z_i^n)+E(Y_i^n)}=\frac{\frac{1}{1-p_i^{\rm{(ret)}}}}{2\frac{1}{1-p_i^{\rm{(ret)}}} + \frac{1}{p_i^{\rm{(vis)}}}} = \frac{p_i^{\rm{(vis)}}}{2p_i^{\rm{(vis)}}+1-p_i^{\rm{(ret)}}} . \label{Eq:pi-k}
\end{equation}
Because $p_i^{\rm{(ret)}}=O(1)$ and $p_i^{\rm{(vis)}}=O(1/N)$, removing higher order terms in Eq.~\eqref{Eq:pi-k} yields
\begin{equation}\label{Eq:pi-k-hat}
\hat\pi_i\approx \frac{p_i^{\rm{(vis)}}}{1-p_i^{\rm{(ret)}}}\propto \frac{d_i^{\rm{(un)}}+d_i^{\rm{(in)}}}{1-\frac{d_i^{\rm{(un)}}}{d_i^{\rm{(un)}}+d_i^{\rm{(out)}}}E\left(\frac{1}{\tilde D^{\rm{(un)}}+D^{\rm{(out)}}}\right)}.
\end{equation}
The proportionality constant is given by imposing that
$\sum_{i=1}^N\hat\pi_i=1$. If the network is undirected, we obtain
$\hat\pi_i\propto d_i^{\rm{(un)}}$, such that $\hat\pi_i$ coincides with the
exact solution used in Eq.~\eqref{Eq:RDSEst}. If the network is fully directed,
i.e., there are no reciprocal edges and $\alpha=1$,
the estimator is proportional to in-directed degree $d_i^{\rm{(in)}}$.

\subsection{Estimating Selection Probabilities From Out-degrees}\label{SubSec:DirectedDegrees}

A common situation in RDS is that only the out-degrees (i.e.,
$d_i^{\rm{(un)}} + d_i^{\rm{(out)}}$) of respondents are recorded. Then, the estimator of the selection probabilities given by Eq.~\eqref{Eq:pi-k-hat} can not be directly used. To cope with this situation, we estimate the number of undirected, in-directed, and out-directed edges from the observed out-degrees and substitute the estimators $(\hat d^{\rm{(un)}}_i,\hat d^{\rm{(in)}}_i,\hat d^{\rm{(out)}}_i)$ in Eq.~\eqref{Eq:pi-k-hat}.

Assume that we have observed the out-degree $d_i^{\rm{(un)}} + d_i^{\rm{(out)}}$ of vertex $i$.
We estimate $d^{\rm{(un)}}_i$ and $d^{\rm{(out)}}_i$ by their expected proportions of the out-degree, and the in-directed degree by its expectation, as follows:
\begin{equation}
\left\{\begin{array}{rcl}
\hat d^{\rm{(un)}}_i &=& \frac{E(D^{\rm{(un)}})}{E(D^{\rm{(un)}})+E(D^{\rm{(out)}})}
\left(d_i^{\rm{(un)}} + d_i^{\rm{(out)}}\right),\\
\hat d^{\rm{(out)}}_i &=& \frac{E(D^{\rm{(out)}})}{E(D^{\rm{(un)}})+E(D^{\rm{(out)}})}
\left(d_i^{\rm{(un)}} + d_i^{\rm{(out)}}\right),\\
\hat d_i^{\rm{(in)}} &=& E(D^{\rm{(in)}}).
\end{array}\right.\label{Eq:EstimateDegrees}
\end{equation}

The expectations used in Eq.~\eqref{Eq:EstimateDegrees} rely on the assumption
that we have a random sample from the network, which is not true in this case. A
plausible assumption on the sampled degree distributions is that they are
size-biased. However, our numerical results suggest that a size-biased
distribution for un-directed and/or the in-directed degree makes little
difference, and if any, increases the bias of selection probability estimators.
Therefore, we stay with the estimators given by Eq.~\eqref{Eq:EstimateDegrees}.

When $(\hat d^{\rm{(un)}}_i,\hat d^{\rm{(in)}}_i,\hat d^{\rm{(out)}}_i)$ is substituted in Eq.~\eqref{Eq:pi-k-hat} in place of $(d^{\rm{(un)}}_i, d^{\rm{(in)}}_i, d^{\rm{(out)}}_i)$, $\hat d_i^{\rm{(un)}}/(\hat d_i^{\rm{(un)}}+\hat d_i^{\rm{(out)}})$ in the denominator is a constant. Therefore, the estimator is proportional to $\hat d^{\rm{(un)}}_i+\hat d^{\rm{(in)}}_i$, i.e., equivalent to Eq.~\eqref{Eq:EstPropIndeg} calculated with the estimated degrees.

\subsection{Estimating Network Parameters}\label{SubSec:NetworkParameters}

The estimators of directed degrees in Eq.~\eqref{Eq:EstimateDegrees} rely on
knowing $E(D^{\rm{(un)}})$, $E(D^{\rm{(in)}})$, and $E(D^{\rm{(un)}})$
separately, which are not estimable from a typical RDS sample, where only the
out-degrees $d_i^{\rm{(un)}} + d_i^{\rm{(out)}}$ of respondents are recorded.
Therefore, we need to extend the estimation procedure to handle these unknown
moments. We do so by assuming a model for the network from which we can estimate
the required moments.

Specifically, we assume that the observed network is a realization of a directed
equivalent of the simple $G(N,p=\lambda/(N-1))$ random graph \citep{Erdos}.
Given parameters $\alpha\in[0,1]$ and $\lambda\in[0,N-1]$, each pair of vertices
independently forms an edge with probability $\lambda/(N-1)$, which is
undirected with probability $(1-\alpha)$ and directed with probability $\alpha$.
When the edge is directed, the direction is selected with equal probability. It
follows that $\lambda$ is the expected total degree of a vertex and that
$\alpha$ is the fraction of directed edges as $N\rightarrow\infty$.

If $N$ is large, $D^{\rm{(un)}}$, $D^{\rm{(in)}}$, and $D^{\rm{(out)}}$
approximately follow independent Poisson distributions with parameters
$(1-\alpha)\lambda$, $\alpha\lambda/2$, and $\alpha\lambda/2$, respectively.
Therefore, the out-degree $D^{\rm{(un)}}+D^{\rm{(out)}}$ and the in-degree
$D^{\rm{(un)}}+D^{\rm{(in)}}$ are both Poisson distributed with parameter
$(2-\alpha)\lambda/2$. Consequently, if we estimate $\alpha$ and $\lambda$, we
can estimate the unknown moments by substituting the estimated $\hat\alpha$ and
$\hat\lambda$ in the moments of the (Poissonian) degree distributions.

To find estimators of $\alpha$ and $\lambda$, we again consider the random walk $\bm X=\{X(t)\}$ on the network. Assume that $e_{ij}=1$, $X(t_0)=i$, and $X(t_0+1)=j$, for a large $t_0$. If $X(t_0+2)=i$, an undirected edge between $i$ and $j$ exists, i.e.\ $e_{ij}=e_{ji}=1$, and the random walk leaves vertex $j$ via $e_{ji}$. Because the edge between $i$ and $j$ is either in-directed to $j$ or undirected, the probability that the edge is undirected is equal to the probability that a randomly selected edge among all undirected and in-directed edges is undirected, i.e.,
$(1-\alpha)/(1-\alpha/2)$. If there is an undirected edge between $i$ and $j$
(i.e., $e_{ji}=1$), the random walk leaves $j$ via $e_{ji}$ with probability
$1/(d_j^{\rm{(un)}}+d_j^{\rm{(out)}})$. Thus, the random walk revisits vertex
$i$ at $t_0+2$ under the directed E-R random graph model with probability
\begin{equation}\label{Eq:Prob-of-revisit}
\frac{1-\alpha}{1-\alpha/2}\cdot\frac{1}{d_j^{\rm{(un)}}+d_j^{\rm{(out)}}}.
\end{equation}

Let $M$ be the number of immediate revisits, which is described above, during $l$ consecutive steps. Then, we have $M=\sum_{k=2}^lM_k$, where $M_k=1$ if a revisit occurs in step $k$ and $M_k=0$ otherwise. $M_k$ is Bernoulli distributed, $M_k\sim{\rm Be}\left((1-\alpha)/(1-\alpha/2)\cdot 1/(d_{j_{k-1}}^{\rm{(un)}}+d_{j_{k-1}}^{\rm{(out)}})\right)$, where $j_{k-1}$ is the vertex visited in step $k-1$. We obtain the expected number of immediate revisits as
\begin{equation}\label{Eq:ExpectedRevisits}
E(M) = \frac{1-\alpha}{1-\alpha/2}\sum_{k=1}^{l-1} \frac{1}{d_{j_k}^{\rm{(un)}}+d_{j_k}^{\rm{(out)}}}.
\end{equation}
If $m$ is the observed number of revisits, we set $m=E(M)$ in Eq.~\eqref{Eq:ExpectedRevisits} to obtain the moment estimator
\begin{equation}\label{Eq:AlphaHat}
\hat\alpha=\frac{m-\sum_{k=1}^{l-1} \left(d_{j_k}^{\rm{(un)}}+d_{j_k}^{\rm{(out)}}\right)^{-1}}{m/2-\sum_{k=1}^{l-1} \left(d_{j_k}^{\rm{(un)}}+d_{j_k}^{\rm{(out)}}\right)^{-1}}.
\end{equation}
If the estimated $\hat\alpha < 0$, we force $\hat\alpha=0$.

Given $\hat\alpha$, we estimate $\lambda$ as follows. If $\alpha=0$, the network contains only undirected edges, and the observed out-degree equals the observed undirected degree, which has a size-biased distribution, with $E(\tilde D^{\rm (un)})=\lambda+1$. If $\alpha=1$, the network has only directed edges, and the expected observed out-degree equals the expected number of out-directed edges, $\lambda/2$. By linearly interpolating the expected observed out-degree between $\alpha=0$ and $\alpha=1$, and substituting it with the mean sample out-degree $\bar u$, we obtain $\bar u = \lambda/2 + (1-\alpha)(1+\lambda/2)$, which yields an estimator of $\lambda$ as
\begin{equation}\label{Eq:LambdaHat}
\hat\lambda = \frac{\bar u+\hat\alpha-1}{1-\hat\alpha/2}.
\end{equation}

Using $\hat\alpha$ and $\hat\lambda$, we can estimate the moments of the degree distributions under the random graph model. For example, $E(D^{\rm{(un)}})$ is estimated by $(1-\hat\alpha)\hat\lambda$. By substituting the estimated moments in Eqs.~\eqref{Eq:pi-k-hat} and \eqref{Eq:EstimateDegrees}, we obtain an estimator of the selection probability of vertex $i$ as

\begin{equation}\label{Eq:Pi-kEstDegEstPar}
\hat\pi_i\propto\hat d^{\rm{(un)}}_i+\hat
d^{\rm{(in)}}_i=\frac{1-\hat\alpha}{1-\hat\alpha/2}
(d_i^{\rm{(un)}}+d_i^{\rm{(out)}})+\frac{\hat\alpha\hat\lambda}{2}.
\end{equation}

When $\alpha=0$ is assumed known and used in place of $\hat\alpha$,
the estimator in Eq.~\eqref{Eq:Pi-kEstDegEstPar} is equivalent to
that used in Eq.~\eqref{Eq:RDSEst}. When
$\hat\alpha=\alpha=1$, it is proportional to 1, and thus equivalent to the
sample mean.

\section{SIMULATION SETUP}\label{Sec:Simulation}

We numerically examine the accuracy of our estimation schemes on directed Erd\H{o}s-Renyi graphs, a model of directed power-law networks (i.e., networks with a power-law degree distribution), and a real online MSM social network. We evaluate both the estimated selection probabilities and corresponding estimates of $p_A$. As described in Section \ref{Sec:Introduction}, real directed social networks show a varying fraction of directed edges, corresponding to a diversity of $\alpha$ values. Therefore, $\alpha$ is varied in the model networks. We also vary $\lambda$ and other network parameters.
We study the performance of the estimators described in Section \ref{Sec:Estimation} when the full degree is observed and when only the out-degree is observed, and compare the performance of our estimators to existing estimators. We do not consider RDS estimators that are not based on the random walk framework because they fall outside the scope of this study.

\subsection{Network Models and Empirical Network}\label{SubSec:NetworkModels}

The first model network that we use is a variant of
the simple Erd\H{o}s-Rényi graph with a mixture of undirected and directed edges, as described in Section~\ref{SubSec:NetworkParameters}. We generate the networks with
$\alpha\in\{0.25,0.5,0.75\}$ and $\lambda\in\{5,10,15\}$. We then extract the largest strongly connected component of the generated network, which has $O(N)$ vertices for all combinations of $\alpha$ and $\lambda$.

The directed Erd\H{o}s-Rényi networks have Poisson degree distributions with quickly decaying tails. To mimic heavy-tailed degree distributions present in many empirical networks~\citep{Newman2010}, we also use
a variant of the power-law network model proposed in \citep{Goh2001PRL,Chung2002PNAS,Chung2003PNAS}.
The original algorithm for generating undirected power-law networks presented in \citet{Goh2001PRL} is as follows.

We fix the number of vertices $N$ and expected degree $E(D)$.
Then, we set the weight of vertex $i$ ($1\le i\le N$) to be $w_i=i^{-\tau}$,
where $0\le \tau\le 1$ is a parameter that controls the power-law exponent of the degree distribution. Then, we select a pair of
vertices $i$ and $j$ ($1\le i\neq j\le N$) with probability proportional to $w_iw_j$.
If the two vertices are not yet connected, we connect them by an undirected edge. We repeat the procedure until the network has $E(D)N/2$ edges.
The expected degree of vertex $i$ is proportional to $w_i$, and the degree distribution is given by $p(d)\propto d^{-\gamma}$, where $\gamma=1+\frac{1}{\tau}$ \citep{Goh2001PRL}.

To generate a power-law network in which undirected and directed edges are mixed with a desired fraction, we extend the algorithm as follows.
First, we specify the expected undirected degree $E(D^{\rm{(un)}})$ and generate an undirected network. Second, we define
$w_i^{\rm in} = (\sigma^{\rm in}(i))^{-\tau^{\rm in}}$ ($1\le i\le N$), where $\sigma^{\rm in}$ is a random permutation on 1, $\ldots$, $N$, and $\tau^{\rm in}$ is a parameter that specifies the power-law exponent of the in-directed degree distribution. Similarly, we set
$w_i^{\rm out}= (\sigma^{\rm out}(i))^{-\tau^{\rm out}}$ ($1\le i\le N$).
Third, we select a pair of vertices with probability proportional to $w_i^{\rm in}w_j^{\rm out}$. If $i\neq j$ and there is not yet a directed edge from $j$ to $i$, we place a directed edge from $j$ to $i$. We repeat the procedure until a total of $E(D^{\rm{(in)}})N/2$ edges are placed. It should be noted that $E(D^{\rm{(in)}})=E(D^{\rm{(out)}})$. The in-directed degree distribution is given by
$p(d^{\rm in})\propto (d^{\rm in})^{-\gamma^{\rm in}}$, where $\gamma^{\rm in}=1+\frac{1}{\tau^{\rm in}}$, and similar for the out-directed degree distribution.
Finally, we superpose the obtained undirected network and directed network to make a single graph. If the combined graph is not strongly connected, we discard it and start over. This network is devoid of degree correlation by construction.

In both network models, we vary the probability of a vertex being assigned
property $A$ as proportional to six different combinations of its degree:
in-degree, out-degree, undirected degree, in-directed degree, out-directed
degree, and directed (in- and out-directed) degree. Formally, if $P(\text{vertex
}i\text{ has }A)\propto g(d_i^{\rm{(un)}},d_i^{\rm{(in)}},d_i^{\rm{(out)}})$, we
let $g$ be equal to $(d_i^{\rm{(un)}}+d_i^{\rm{(in)}})$,
$(d_i^{\rm{(un)}}+d_i^{\rm{(out)}})$, $d_i^{\rm{(un)}}$, $d_i^{\rm{(in)}}$,
$d_i^{\rm{(out)}}$, and $(d_i^{\rm{(in)}}+d_i^{\rm{(out)}})$, respectively. We
refer to these as different ways to allocate property $A$. We also examined
the case in which we assigned the property uniformly over all vertices. However,
because the performance of the different estimators is almost the same in this
case, we do not show the results in the following. For all allocations of $A$,
the property is assigned in such a way that the expected proportion of vertices
being assigned $A$ is equal to some fixed value $p$.
Because $A$ is stochastically assigned, the actual proportion $p_A$ of vertices with $A$ will vary between realized allocations.

We also evaluate our estimators on an online MSM social network,
www.qruiser.com, which is the Nordic region's largest community for lesbian,
gay, bisexual, transgender and queer persons~\citep[Dec 2005-Jan
2006;][]{Rybski2009,LuEtal,Lu2012JRSS}. Our dataset consists of 16,082 male
homosexual members and forms a strongly connected component. Because members are
allowed to add any member to their list of contacts without approval of that
member, the resulting network is directed; the fraction of directed edges equals
$\alpha=0.7572$. The in-degree and out-degree distributions are
skewed~\citep{Lu2012JRSS}, and the mean number of edges $\lambda$ is equal to
27.7434. The data set also includes user's profiles, from which  we obtain four
dichotomous properties on which we evaluate estimators of population
proportions: age (born before 1980 or not), county (live in Stockholm or not),
civil status (married or unmarried), and profession (employed or unemployed).

\subsection{Evaluation of Estimators}

We compared the performance of our estimators of the selection probabilities with three other estimators. We refer to our estimator $\{\hat\pi_i\}$ obtained from Eq.~\eqref{Eq:pi-k-hat} as $\{\hat\pi_i^{\rm{(ren)}}\}$ (ren stands for renewal). The other estimators are the uniform stationary distribution $\{\hat\pi_i^{\rm{(uni)}}\}$, where $\hat\pi_i^{\rm{(uni)}} = 1/N$ for all $i$, the selection probabilities proportional to the out-degree $\{\hat\pi_i^{\rm{(outdeg)}}\}$, on which Eq.~\eqref{Eq:RDSEst} is based, where $\hat\pi_i^{\rm{(outdeg)}}\propto d_i^{\rm{(un)}}+d_i^{\rm{(out)}}$, and the stationary distribution obtained from Eq.~\eqref{Eq:EstPropIndeg} $\{\hat\pi_i^{\rm{(indeg)}}\}$, i.e., proportional to the in-degree. In the following, we suppress the $\{\}$ notation.

To assess the performance of an estimator we first
calculated the estimated selection probabilities $\hat\pi_i$ for one of the four estimators
and the true stationary distribution $\pi_i$ at all the vertices in the given network. Then, we calculated their \emph{total variation distance} defined by
\begin{equation}
D_{TV}=\frac{1}{2}\sum_{i=1}^N|\hat \pi_i-\pi_i|
\end{equation}
\citep{Levin}. The stationary distribution $\pi_i$ was obtained using the power method~\citep{Langville2006book} with an accuracy of $10^{-10}$ in terms of the total variation distance for the two distributions given in the successive two steps of the power iteration.

For $\hat\pi^{\rm{(ren)}}$, we considered three variants depending on the information available from observed degree and knowledge of the moments of the degree distributions. When the full degree
$(d_i^{\rm{(un)}},d_i^{\rm{(in)}},d_i^{\rm{(out)}})$ is observed, we used
Eq.~\eqref{Eq:pi-k-hat} to calculate $\hat\pi^{\rm{(ren)}}$, where
$E\left(1/(\tilde D^{\rm{(un)}}+D^{\rm{(out)}})\right)$ is estimated by the mean
of the inverse sample out-degrees. We denote the corresponding estimator with
$\hat\pi^{\rm{(ren)}}_{\rm{f.d.}}$, where f.d.\ stands for ``full degree''. When
only the out-degree is observed and the moments of the degree distributions are
known, we used Eq.~\eqref{Eq:EstimateDegrees}. This case is only evaluated for
the directed Erd\H{o}s-Rényi graphs, and the corresponding estimator is denoted
by $\hat\pi^{\rm{(ren)}}_{\alpha,\lambda}$. If only the out-degree is observed
and the moments of the degree distributions are unknown, we used
Eqs.~\eqref{Eq:AlphaHat}, \eqref{Eq:LambdaHat}, and \eqref{Eq:Pi-kEstDegEstPar}, and the estimator is denoted $\hat\pi^{\rm{(ren)}}$.

We sampled from each generated network by means of a random walk starting from a randomly selected vertex. In the random walk, we collect the degree of the visited nodes
and also check whether they have property $A$ or not. We estimated the population proportion $p_A$ from the sample by replacing $\pi$ in Eq.~\eqref{Eq:GenProbEst} by
either $\hat\pi^{\rm{(uni)}}$, $\hat\pi^{\rm{(outdeg)}}$, $\hat\pi^{\rm{(indeg)}}$, or any of the variants of $\hat\pi^{\rm{(ren)}}$, yielding estimates
$\hat p_A^{\rm{(uni)}}$, $\hat p_A^{\rm{(outdeg)}}$, $\hat p_A^{\rm{(indeg)}}$, or $\hat p_A^{\rm{(ren)}}$, respectively. The sample size is denoted by $s$.

\section{NUMERICAL RESULTS}\label{Sec:Results}

\subsection{Directed Erd\H{o}s-Renyi Graphs}\label{SubSec:ResultsER}

In Table~\ref{Tab:ERObserveAll}, we show the mean of the total variation distance $D_{TV}$ between the true stationary distribution and $\hat\pi^{\rm{(uni)}}$, $\hat\pi^{\rm{(outdeg)}}$, $\hat\pi^{\rm{(indeg)}}$, and $\hat\pi^{\rm{(ren)}}_{\rm{f.d.}}$, calculated on the basis of 1000 realizations of the largest strongly connected component of
the directed random graph having $N=1000$ vertices. Because the standard deviation of $D_{TV}$ is similar between the estimators, we show an average over the four
estimators. The sample size $s$ used in $\hat\pi^{\rm{(ren)}}_{\rm{f.d.}}$ is 500. We also tried $s=200$, which gave similar results. The $D_{TV}$ value of
$\hat\pi^{\rm{(indeg)}}$ and $\hat\pi^{\rm{(ren)}}_{\rm{f.d.}}$ is much smaller than that of $\hat\pi^{\rm{(uni)}}$ and
$\hat\pi^{\rm{(outdeg)}}$ for all values of
$\alpha$ and $\lambda$. Furthermore $\hat\pi^{\rm{(ren)}}_{\rm{f.d}}$ always gives smaller $D_{TV}$ than $\pi^{\rm{(indeg)}}$ although the two values are similar for many
combinations of the parameters.

\begin{table}
\setlength{\tabcolsep}{4.5pt}
\centering
\caption{Mean and average s.d.\ of $D_{TV}$ for the directed random graph when $(d_i^{\rm{(un)}},d_i^{\rm{(in)}},d_i^{\rm{(out)}})$ is observed and moments of the degree
distributions are known. The lowest $D_{TV}$ value marked
in boldface. We set $N=1000$. }\label{Tab:ERObserveAll}
\subfloat[$\alpha=0.1$]{
\begin{tabular}{@{}cccccc@{}}
\hline\hline
$\lambda$ & $\hat\pi^{\rm{(uni)}}$ & $\hat\pi^{\rm{(outdeg)}}$ & $\hat\pi^{\rm{(indeg)}}$ & $\hat\pi^{\rm{(ren)}}_{\rm{f.d}}$ & s.d.\\
\hline
5 & 0.185 & 0.074 & 0.042 & \textbf{0.041} & 0.004\\
10 & 0.131 & 0.045 & 0.017 & \textbf{0.016} & 0.002\\
15 & 0.106 & 0.036 & \textbf{0.010} & \textbf{0.010} & 0.001\\
\end{tabular}}\quad
\subfloat[$\alpha=0.25$]{
\begin{tabular}{@{}ccccc@{}}
\hline\hline
$\hat\pi^{\rm{(uni)}}$ & $\hat\pi^{\rm{(outdeg)}}$ & $\hat\pi^{\rm{(indeg)}}$ & $\hat\pi^{\rm{(ren)}}_{\rm{f.d}}$ & s.d.\\
\hline
 0.203   & 0.134    & 0.077    & \textbf{0.075} & 0.005\\
0.140    &0.081    &0.031    &\textbf{0.030} & 0.002\\
 0.112 &   0.063  &  \textbf{0.019}  &  \textbf{0.019} & 0.002\\
\end{tabular}}\\
\subfloat[$\alpha=0.5$]{
\begin{tabular}{@{}cccccc@{}}
\hline\hline
$\lambda$ & $\hat\pi^{\rm{(uni)}}$ & $\hat\pi^{\rm{(outdeg)}}$ & $\hat\pi^{\rm{(indeg)}}$ & $\hat\pi^{\rm{(ren)}}_{\rm{f.d}}$ & s.d.\\
\hline
5 & 0.247 & 0.225 & 0.138 & \textbf{0.133} & 0.009\\
10 & 0.160 & 0.136 & 0.056 & \textbf{0.055} & 0.004\\
15 & 0.126 & 0.105 & 0.034 & \textbf{0.033} & 0.002\\
\end{tabular}}\quad
\subfloat[$\alpha=0.75$]{
\begin{tabular}{@{}ccccc@{}}
\hline\hline
$\hat\pi^{\rm{(uni)}}$ & $\hat\pi^{\rm{(outdeg)}}$ & $\hat\pi^{\rm{(indeg)}}$ & $\hat\pi^{\rm{(ren)}}_{\rm{f.d}}$ & s.d.\\
\hline
0.303 & 0.319 & 0.207 & \textbf{0.201} & 0.014\\
0.188  &   0.201  &  0.090  &  \textbf{0.088} & 0.005\\
0.144 &   0.156 &   \textbf{0.055} &   \textbf{0.055} & 0.003\\
\end{tabular}}
\end{table}

In Table~\ref{Tab:ERObserveOutdegreeEstimatedParameters}, we show the
mean and average s.d.\ of $D_{TV}$ when the out-degree, i.e.\
$d_i^{\rm{(un)}}+d_i^{\rm{(out)}}$, is observed but the individual $d_i^{\rm{(un)}}$ and
$d_i^{\rm{(out)}}$ values are not. The assumptions underlying the network generation are the same as those for Table~\ref{Tab:ERObserveAll}, and the sample size $s$ is
equal to 500.
Here we consider two cases. In the first case, the moments of the degree distribution are known, and we use the estimator
$\hat\pi^{\rm{(ren)}}_{\alpha,\lambda}$. In the second case, they are not known, and we use $\hat\pi^{\rm{(ren)}}$. Results for $\hat\pi^{\rm{(indeg)}}$
are not shown in Table~\ref{Tab:ERObserveOutdegreeEstimatedParameters}
because in-degree is not observed.
Table~\ref{Tab:ERObserveOutdegreeEstimatedParameters} indicates that
$D_{TV}$ for $\hat\pi^{\rm{(ren)}}$
is smaller than that for $\hat\pi^{\rm{(uni)}}$ and $\hat\pi^{\rm{(outdeg)}}$
when $\alpha$ is 0.5 and 0.75.
When
$\alpha=0.75$, $\hat\pi^{\rm{(outdeg)}}$ yields the largest
$D_{TV}$. For $\alpha=0.1$ and 0.25, $\hat\pi^{\rm{(ren)}}$ and
$\hat\pi^{\rm{(outdeg)}}$ yield similar results. For all parameter values
$\hat\pi^{\rm{(ren)}}_{\alpha,\lambda}$ slightly outperforms
$\hat\pi^{\rm{(ren)}}$. We tried $s=200$ (not shown) which gave similar s.d. for
$\hat\pi^{\rm{(ren)}}_{\alpha,\lambda}$, and similarly for
$\hat\pi^{\rm{(ren)}}$, except for $\alpha=0.1$,
where, for example, $\lambda=15$ yielded the s.d. values of 0.0039 and 0.0073 for $s=500$ and $s=200$, respectively.

\begin{table}
\setlength{\tabcolsep}{4.5pt}
\centering
\caption{Mean and average s.d.\ of $D_{TV}$ for the directed random graph when $d_i^{\rm{(un)}}+d_i^{\rm{(out)}}$ is observed.
We set $N=1000$.}\label{Tab:ERObserveOutdegreeEstimatedParameters}
\subfloat[$\alpha=0.1$]{
\begin{tabular}{@{}cccccc@{}}
\hline\hline
$\lambda$ & $\hat\pi^{\rm{(uni)}}$ & $\hat\pi^{\rm{(outdeg)}}$ & $\hat\pi^{\rm{(ren)}}_{\alpha,\lambda}$ & $\hat\pi^{\rm{(ren)}}$ & s.d.\\
\hline
5 & 0.185 & \textbf{0.074} & \textbf{0.074} & 0.075 & 0.004\\
10 & 0.131 & \textbf{0.045} & \textbf{0.045} & 0.047 & 0.003\\
15 & 0.106 & 0.036 & \textbf{0.035} &  0.037 & 0.002\\
\end{tabular}}\quad
\subfloat[$\alpha=0.25$]{
\begin{tabular}{@{}ccccc@{}}
\hline\hline
$\hat\pi^{\rm{(uni)}}$ & $\hat\pi^{\rm{(outdeg)}}$ & $\hat\pi^{\rm{(ren)}}_{\alpha,\lambda}$ & $\hat\pi^{\rm{(ren)}}$ & s.d.\\
\hline
 0.203 & 0.135 & \textbf{0.132} & 0.133 & 0.006\\
0.140 & 0.081 & \textbf{0.079} & 0.080 & 0.003\\
0.112 & 0.063 & \textbf{0.061} & 0.063 & 0.002\\
\end{tabular}}\\
\subfloat[$\alpha=0.5$]{
\begin{tabular}{@{}cccccc@{}}
\hline\hline
$\lambda$ & $\hat\pi^{\rm{(uni)}}$ & $\hat\pi^{\rm{(outdeg)}}$ & $\hat\pi^{\rm{(ren)}}_{\alpha,\lambda}$ & $\hat\pi^{\rm{(ren)}}$ & s.d.\\
\hline
5 & 0.246 & 0.225 & \textbf{0.214} & 0.215 & 0.010\\
10 & 0.160 & 0.136 & \textbf{0.127} & 0.128 & 0.004\\
15 & 0.125 & 0.105 & \textbf{0.098} & 0.099 & 0.003\\
\end{tabular}}\quad
\subfloat[$\alpha=0.75$]{
\begin{tabular}{@{}ccccc@{}}
\hline\hline
$\hat\pi^{\rm{(uni)}}$ & $\hat\pi^{\rm{(outdeg)}}$ & $\hat\pi^{\rm{(ren)}}_{\alpha,\lambda}$ & $\hat\pi^{\rm{(ren)}}$ & s.d.\\
\hline
0.303 & 0.318 & \textbf{0.294} & 0.295 & 0.014\\
0.188 & 0.201 & \textbf{0.177} &  0.178 & 0.006\\
0.144 & 0.156 & \textbf{0.135} &  0.135 & 0.004\\
\end{tabular}}
\end{table}

\begin{figure}
\centering
\caption{Deviations of estimated $\hat p_A$ from true value in the directed
Erd\H{o}s-Rényi graphs with $N=1000$, $\alpha=0.75$, $\lambda=10$,
$p=0.5$, and $s=500$. Each group of boxplots corresponds to $\hat
p_{A_{\rm{f.d.}}}^{\rm{(ren)}}$, $\hat p_A^{\rm{(indeg)}}$, $\hat
p_A^{\rm{(ren)}}$, $\hat p_{A_{\alpha,\lambda}}^{\rm{(ren)}}$, $\hat
p_A^{\rm{(outdeg)}}$, and $\hat p_A^{\rm{(uni)}}$ for one allocation of the
individual property $A$. The abbreviations for the allocations corresponds
to the function $g$, i.e., In-deg. equals $(d_i^{\rm{(un)}}+d_i^{\rm{(in)}})$,
Out-deg. $(d_i^{\rm{(un)}}+d_i^{\rm{(out)}})$, Undir. $d_i^{\rm{(un)}}$, In-dir.
$d_i^{\rm{(in)}}$, Out-dir. $d_i^{\rm{(out)}}$, and Dir.
$(d_i^{\rm{(in)}}+d_i^{\rm{(out)}})$.}
\includegraphics[width=13cm]{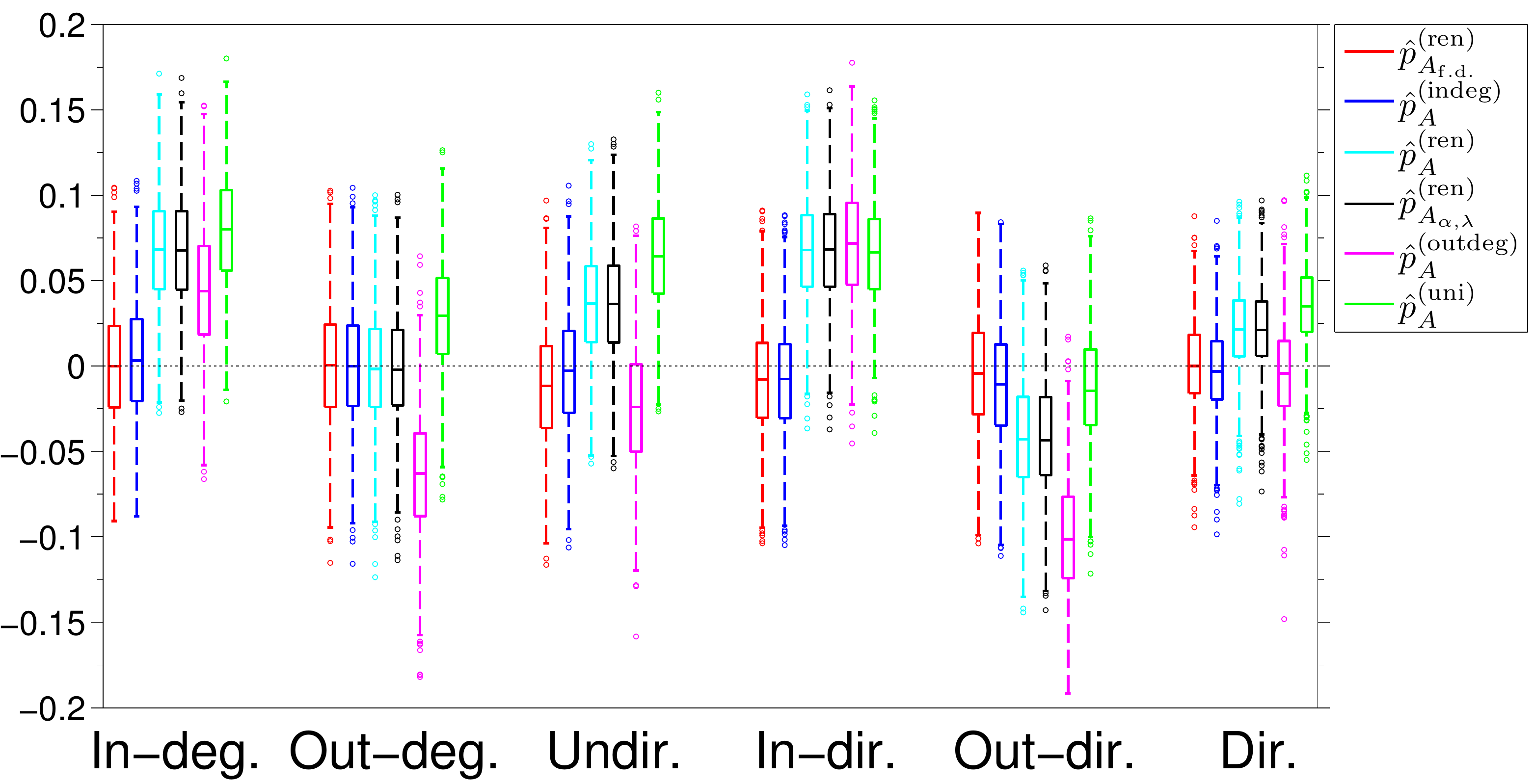}\label{Fig:ERProportions}
\end{figure}

To compare estimated $p_A$, we generated 1000 networks for each combination of
the parameters $\alpha\in\{0.25,0.5,0.75\}$ and $\lambda=10$. On
each of these networks we in turn allocate the property $A$ in each of the six ways described in Section~\ref{SubSec:NetworkModels}. The probability of a vertex having
$A$ is denoted by $p\in\{0.2,0.5\}$. For each network and allocation, we simulate a random walk with length
$s\in\{200,500\}$ and calculate the differences between
estimated proportions of the population with property $A$ and the actual proportion of vertices with $A$. In
Figure~\ref{Fig:ERProportions}, results for $\alpha=0.75$, $p=0.5$, and $s=500$
are shown. The six groups of four boxplots
correspond to the six different ways of allocating $A$ (see
Section~\ref{SubSec:NetworkModels}). The six boxplots in each
group correspond to $\hat
p_{A_{\rm{f.d.}}}^{\rm{(ren)}}$, $\hat p_A^{\rm{(indeg)}}$, $\hat
p_A^{\rm{(ren)}}$, $\hat p_{A_{\alpha,\lambda}}^{\rm{(ren)}}$, $\hat
p_A^{\rm{(outdeg)}}$, and $\hat p_A^{\rm{(uni)}}$.

We see that the bias of $\hat p_{A_{\rm{f.d}}}^{\rm{(ren)}}$ and $\hat p_A^{\rm{(indeg)}}$
is small for all allocations, as to be expected. For the estimators utilizing the out-degree, $\hat
p_A^{\rm{(ren)}}$, $\hat p_{A_{\alpha,\lambda}}^{\rm{(ren)}}$, and $\hat
p_A^{\rm{(outdeg)}}$, Figure~\ref{Fig:ERProportions} indicates that the choice of how to
allocate $A$ has a significant impact on the performance of estimators. When $A$ is allocated proportional to the
out-degree (Out-deg. in Fig.~\ref{Fig:ERProportions}),  $\hat p_A^{\rm{(ren)}}$ and
$\hat p_{A_{\alpha,\lambda}}^{\rm{(ren)}}$ yields the most accurate result, and when $A$ is
allocated proportional to the number of directed edges (Dir. in Fig.~\ref{Fig:ERProportions}), $\hat
p_A^{\rm{(outdeg)}}$ is most accurate; this is true for almost all parameter combinations. In general, the bias and
variance increase with both $\alpha$ and $p$ for all estimators, and a small $s$ results in an increased variance, as to
be expected. In the Supplementary material, these findings are further illustrated by numerical results with
$(\alpha,p,s)$ equal to $(0.5,0.2,500)$, $(0.25,0.5,500)$, and $(0.75,0.5,200)$.

\subsection{Networks With Power-law Degree Distributions}

To generate power-law networks, we set the expected total number of edges for
each node to 16, while we set the expected number of undirected and directed
edges equal to $(E(D^{\rm{(un)}}), E(D^{\rm{(in)}}+D^{\rm{(out)}})) = (12, 4), (8, 8)$, and $(4, 12)$. The three cases
yield $\alpha=0.25$, 0.5, and 0.75, respectively. For each combination of the parameters, we generate 1000 networks of
size $N=1000$ and calculate the mean of the $D_{TV}$. We also calculate the s.d., which is of magnitude $10^{-3}$ and
therefore not shown. The sample size $s$ is set to 200 and 500.

\begin{figure}
\centering
\caption{Average $D_{TV}$ between the true stationary distribution and $\hat\pi^{\rm{(ren)}}_{\rm{f.d.}}$,
$\hat\pi^{\rm{(indeg)}}$, $\hat\pi^{\rm{(ren)}}$,
$\hat\pi^{\rm{(ren)}}_{\alpha,\lambda}$, $\hat\pi^{\rm{(outdeg)}}$, and $\hat\pi^{\rm{(uni)}}$ in the power-law networks
with $N=1000$, $\alpha$ equal to a) 0.25, b) 0.5, and c) 0.75, and $s=500$.}\label{Fig:PL}
\subfloat[$\alpha=0.25$]{
\includegraphics[width=4.8cm]{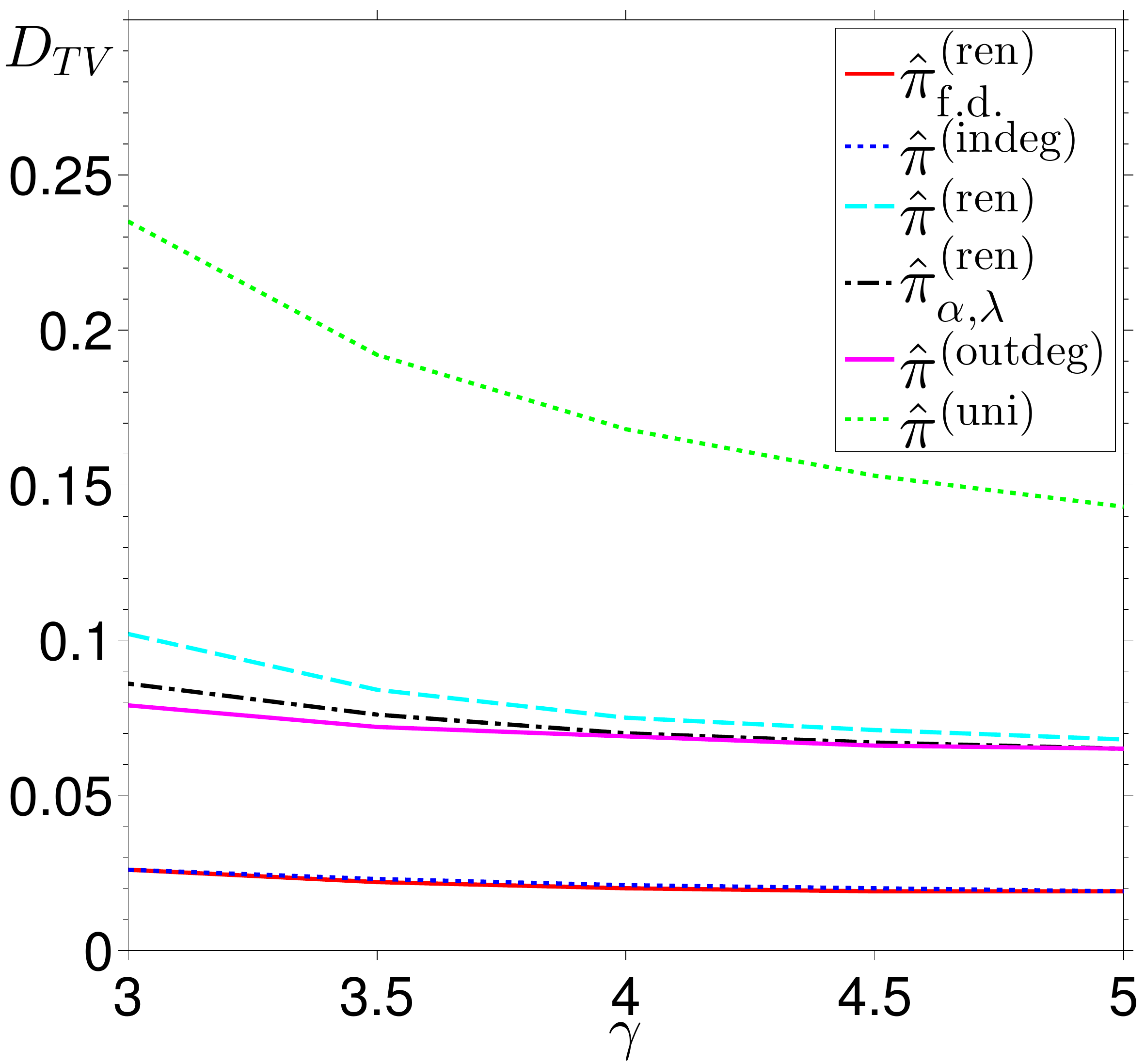}
}
\subfloat[$\alpha=0.5$]{
\includegraphics[width=4.8cm]{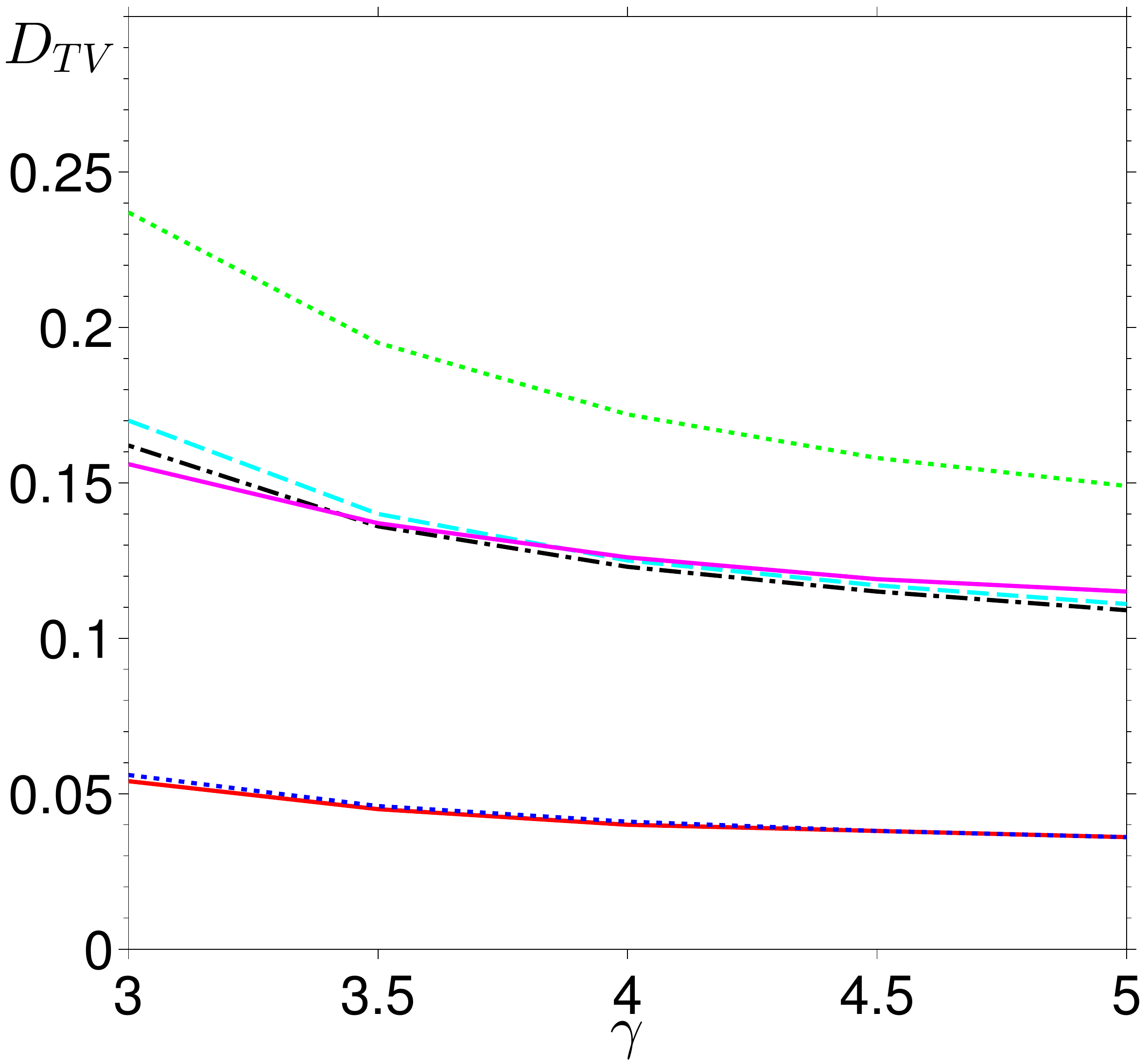}
}
\subfloat[$\alpha=0.75$]{
\includegraphics[width=4.8cm]{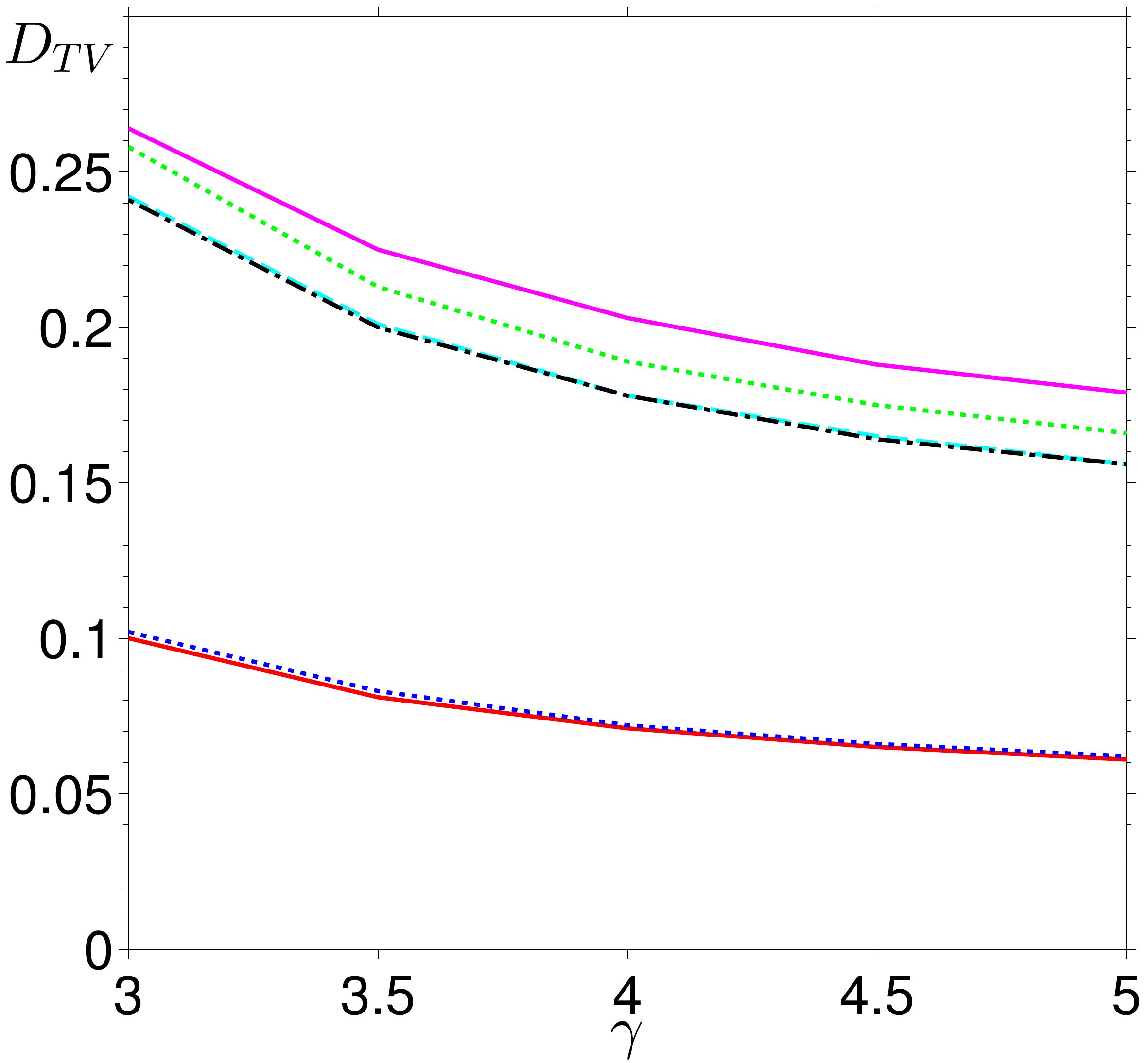}
}
\end{figure}

The average $D_{TV}$ values for $\hat\pi^{\rm{(ren)}}_{\rm{f.d.}}$, $\hat\pi^{\rm{(indeg)}}$, $\hat\pi^{\rm{(ren)}}$,
$\hat\pi^{\rm{(ren)}}_{\alpha,\lambda}$, $\hat\pi^{\rm{(outdeg)}}$, and $\hat\pi^{\rm{(uni)}}$ are shown in
Figure~\ref{Fig:PL} for various $\alpha$ and $\gamma$ values. Figure~\ref{Fig:PL} suggests that
$\hat\pi^{\rm{(ren)}}_{\rm{f.d}}$ and $\hat\pi^{\rm{(indeg)}}$ are the most accurate among the four estimators, with
$\hat\pi^{\rm{(ren)}}_{\rm{f.d}}$ being slightly better. When $\alpha=0.25$ and 0.5,
$\hat\pi^{\rm{(ren)}}_{\alpha,\lambda}$ has a lower mean $D_{TV}$ than $\hat\pi^{\rm{(ren)}}$, but this difference is
not seen when $\alpha=0.75$. $\hat\pi^{\rm{(outdeg)}}$ performs better than $\hat\pi^{\rm{(ren)}}$ for all values of
$\gamma$ when $\alpha=0.25$, and the opposite result holds true when $\alpha=0.75$.

\begin{figure}
\centering
\caption{Deviations of estimated $p_A$ from the true population proportion in the power-law networks for $\gamma=3$,
$E(D^{\rm{(un)}})=4$, $E(D^{\rm{(in)}}+D^{\rm{(out)}})=12$, $p=0.2$, and $s=500$. Each group of boxplots corresponds to
$\hat p_{A_{\rm{f.d.}}}^{\rm{(ren)}}$, $\hat p_A^{\rm{(indeg)}}$, $\hat p_A^{\rm{(ren)}}$,
$\hat p_{A_{\alpha,\lambda}}^{\rm{(ren)}}$, $\hat p_A^{\rm{(outdeg)}}$, and $\hat p_A^{\rm{(uni)}}$, for one allocation
of $A$.}
\includegraphics[width=13cm]{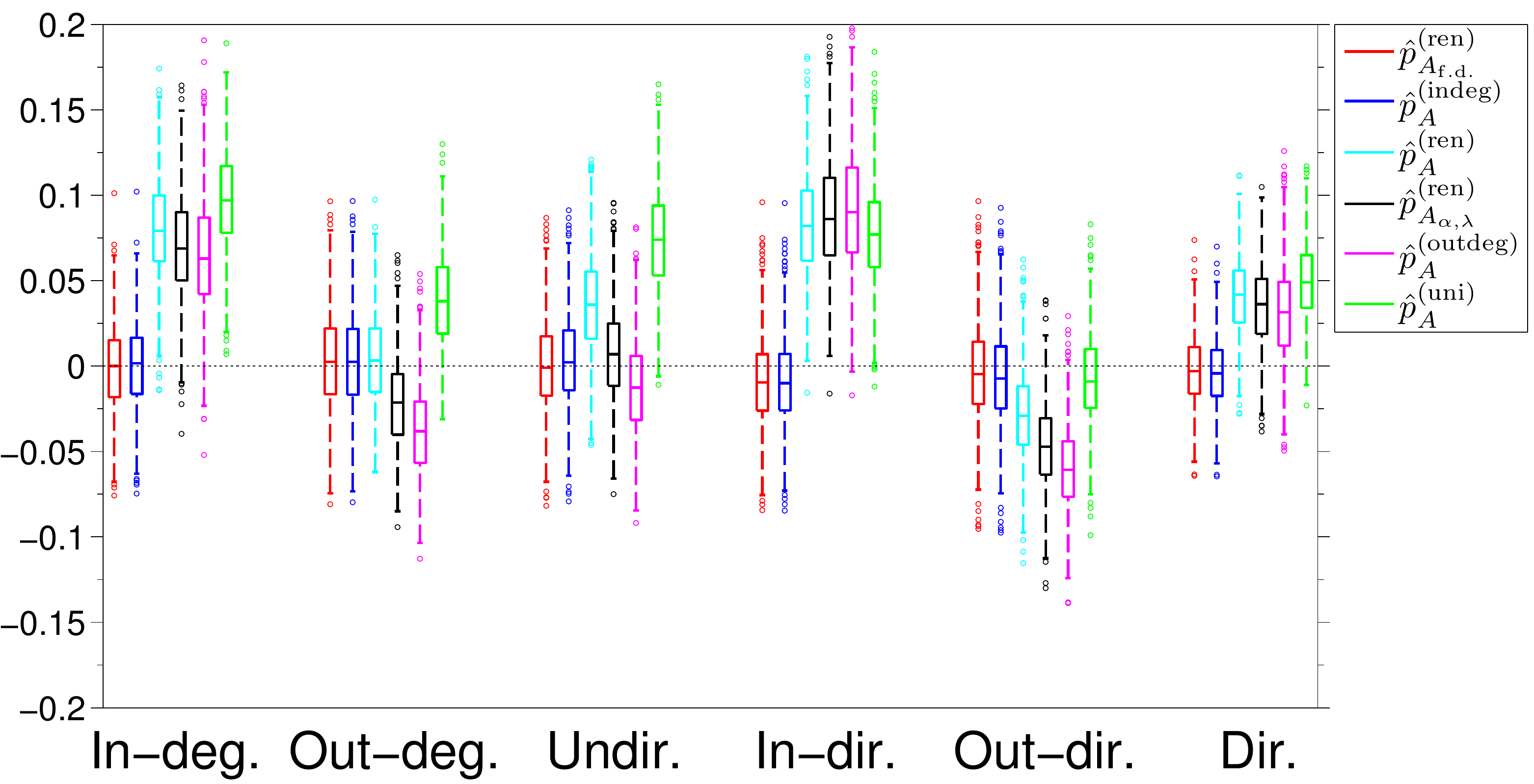}\label{Fig:PLProp}
\end{figure}

In Figure~\ref{Fig:PLProp}, the results for $\hat p_{A_{\rm{f.d.}}}^{\rm{(ren)}}$, $\hat p_A^{\rm{(indeg)}}$,
$\hat p_A^{\rm{(ren)}}$, $\hat p_{A_{\alpha,\lambda}}^{\rm{(ren)}}$, $\hat p_A^{\rm{(outdeg)}}$, and
$\hat p_A^{\rm{(uni)}}$ when $\gamma=3$, $E(D^{\rm{(un)}})=4$, $E(D^{\rm{(in)}}+D^{\rm{(out)}})=12$, $p=0.2$, and
$s=500$ are shown. The figure indicates that $\hat p_{A_{\rm{f.d.}}}^{\rm{(ren)}}$ and $\hat p_A^{\rm{(indeg)}}$ have
small bias across different allocations of $A$. In contrast, the magnitude of the bias of $\hat p_A^{\rm{(ren)}}$,
$\hat p_{A_{\alpha,\lambda}}^{\rm{(ren)}}$, and $\hat p_A^{\rm{(outdeg)}}$ depends on the allocation type; $\hat
p_A^{\rm{(ren)}}$ has the smallest bias when $A$ is allocated proportional to the undirected degree, and $\hat
p_{A_{\alpha,\lambda}}^{\rm{(ren)}}$ and $\hat p_A^{\rm{(outdeg)}}$ when $A$ is allocated proportional to the
out-degree. Their relative performance is hard to assess for other allocations. In general, a large fraction of directed
edges, small $\gamma$, and large $p$ increase bias and variance, and variance of course decreases with $s$. The
Supplementary material contains numerical results
for $(\gamma,E(D^{\rm{(un)}}),E(D^{\rm{(in)}}+D^{\rm{(out)}}),p,s)=(4.5,4,12,0.2,500)$, $(4.5,4,12,0.5,500)$,
$(4.5,12,4,0.5,500)$, and $(3,4,12,0.2,200)$ to further support these results.

\subsection{Online MSM Network}

For the Qruiser online MSM network, we first evaluate $\hat\pi^{\rm{(uni)}}$, $\hat\pi^{\rm{(outdeg)}}$, $\hat\pi^{\rm{(indeg)}}$, $\hat\pi^{\rm{(ren)}}_{\rm{f.d.}}$, and
$\hat\pi^{\rm{(ren)}}$. The results are shown in Table~\ref{Tab:Qruiser}. Note that
$\hat\pi^{\rm{(ren)}}_{\alpha,\lambda}$ is not evaluated because $\alpha$ and $\lambda$ are not known beforehand. For
$\hat\pi^{\rm{(uni)}}$, $\hat\pi^{\rm{(outdeg)}}$, and $\hat\pi^{\rm{(indeg)}}$,
$D_{TV}$ to the true selection probabilities is exactly calculated. For $\hat\pi^{\rm{(ren)}}_{\rm{f.d.}}$ and
$\hat\pi^{\rm{(ren)}}$, we
show the mean and s.d.\ of
$D_{TV}$ on the basis of 1000 samples of size $500$. We see that
$\hat\pi^{\rm{(ren)}}_{\rm{f.d.}}$ has smaller $D_{TV}$ than
$\hat\pi^{\rm{(indeg)}}$, and that the mean $D_{TV}$ of $\hat\pi^{\rm{(ren)}}$
is smaller than that of $\hat\pi^{\rm{(uni)}}$ and $\hat\pi^{\rm{(outdeg)}}$.

\begin{table}
\centering
\caption{$D_{TV}$ between the true stationary distribution and $\hat\pi^{\rm{(uni)}}$, $\hat\pi^{\rm{(outdeg)}}$, $\hat\pi^{\rm{(indeg)}}$, $\hat\pi^{\rm{(ren)}}_{\rm{f.d.}}$ and $\hat\pi^{\rm{(ren)}}$. S.d. is shown in the second row, but only applies to $\hat\pi^{\rm{(ren)}}_{\rm{f.d.}}$ and $\hat\pi^{\rm{(ren)}}$.}\label{Tab:Qruiser}
\begin{tabular}{@{}cccccc@{}}
\hline\hline
$\hat\pi^{\rm{(ren)}}_{\rm{f.d.}}$ & $\hat\pi^{\rm{(indeg)}}$ &
$\hat\pi^{\rm{(ren)}}$ & $\hat\pi^{\rm{(outdeg)}}$ & $\hat\pi^{\rm{(uni)}}$ \\
\hline
      0.2198 & 0.2248 & 0.4057 & 0.4290 & 0.4484 \\
 0.0004 & - & 0.0048 & -
\end{tabular}
\end{table}

\begin{figure}
\centering
\caption{Estimates of population proportions in the Qruiser network for a) age,
b) civil status, c) county, and d) profession. Each figure shows $\hat
p_{A_{\rm{f.d.}}}^{\rm{(ren)}}$, $\hat
p_A^{\rm{(indeg)}}$, $\hat
p_A^{\rm{(ren)}}$, $\hat
p_A^{\rm{(outdeg)}}$, and $\hat
p_A^{\rm{(uni)}}$. The  true population proportions are shown by the dashed
lines and are equal to
0.77, 0.40, 0.39, and 0.38 for age, civil status, county, and profession,
respectively.}\label{Fig:Qruiser}
\subfloat[]{
\includegraphics[width=3.5cm]{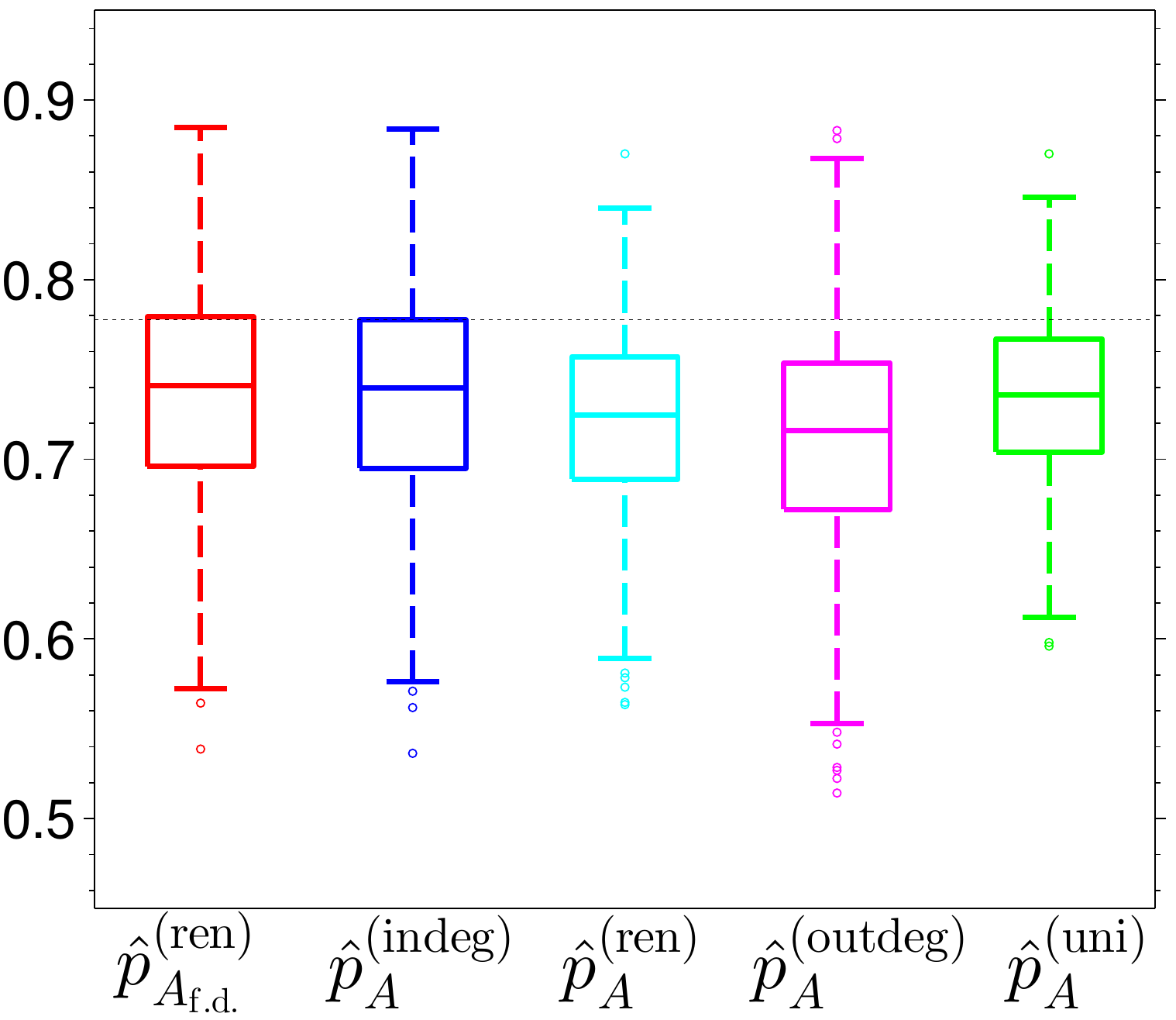}
}
\subfloat[]{
\includegraphics[width=3.5cm]{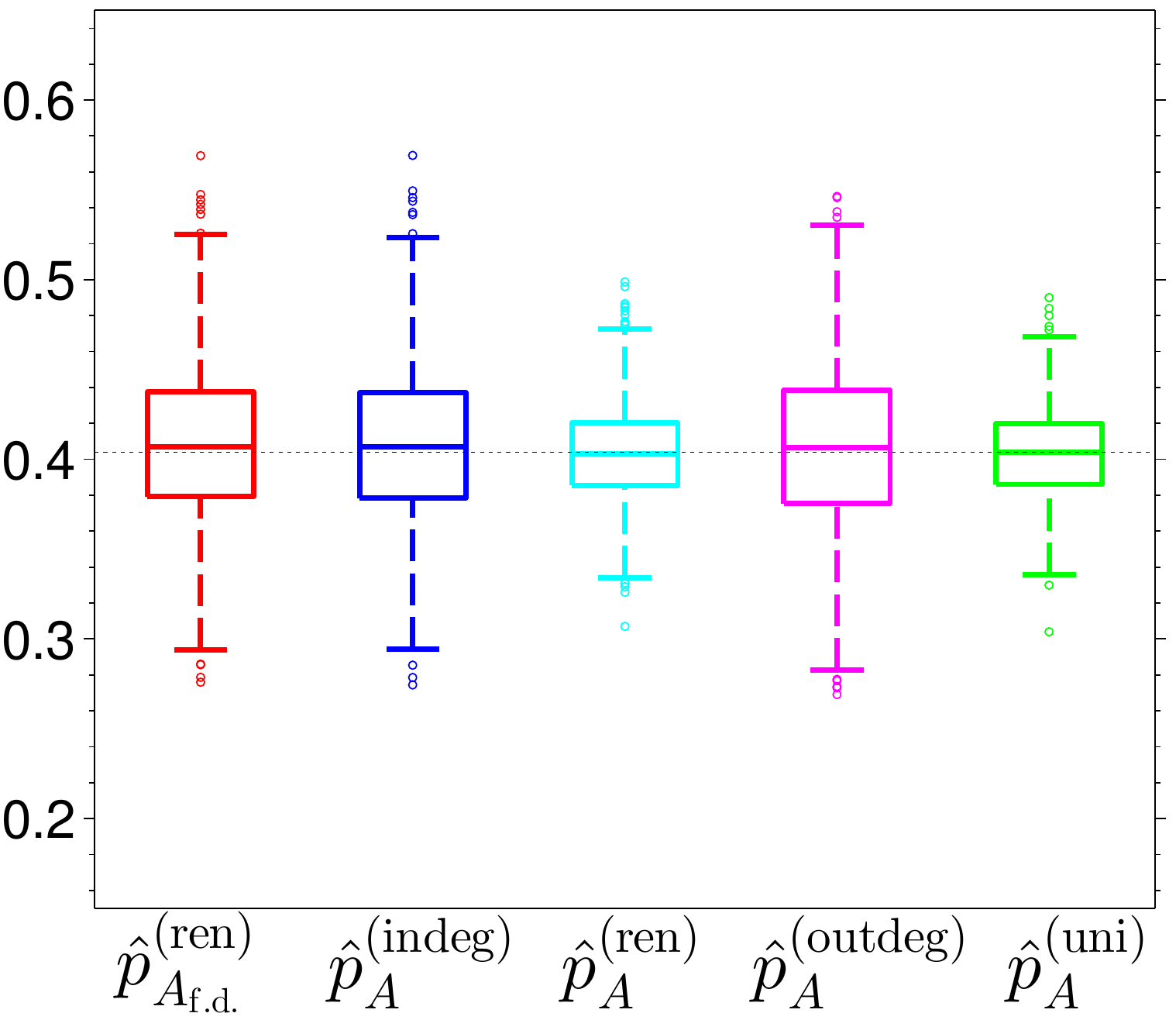}
}
\subfloat[]{
\includegraphics[width=3.5cm]{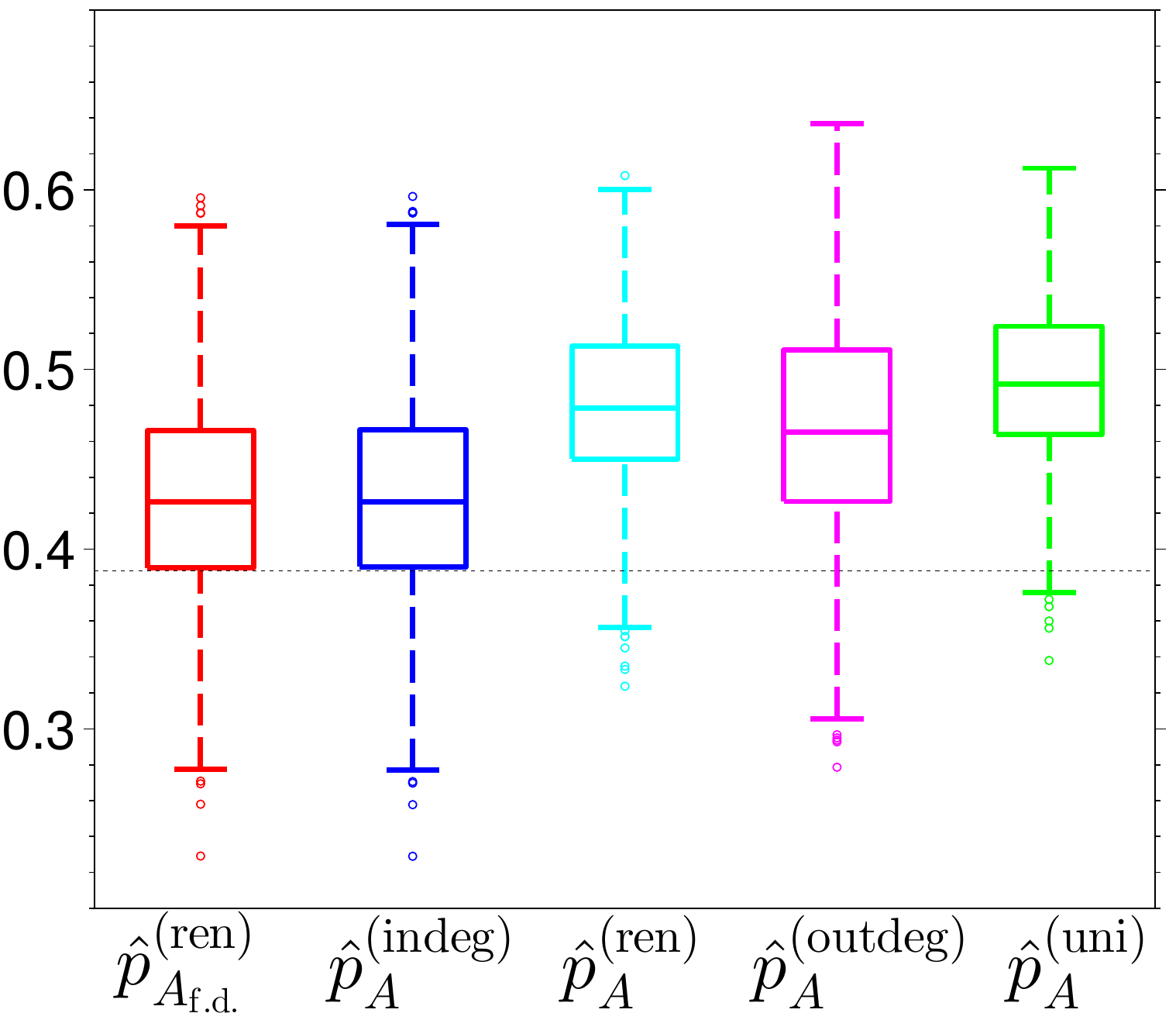}
}
\subfloat[]{
\includegraphics[width=3.5cm]{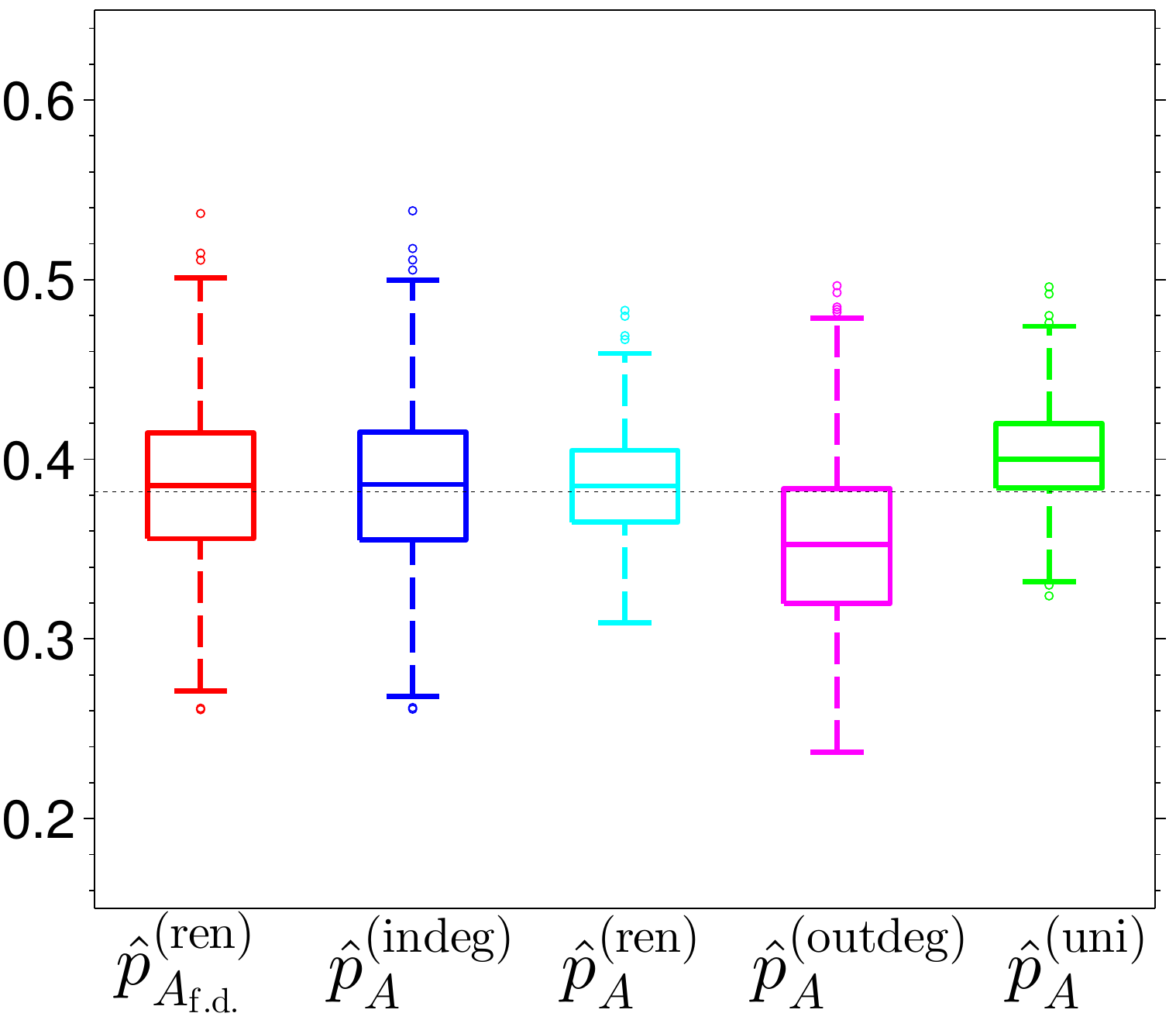}
}
\end{figure}

In Figure~\ref{Fig:Qruiser}, we show estimates of the population
proportions of the age, county, civil status, and profession properties. The
true population proportions are shown by the dashed lines. The
sample size is 500. Figure~\ref{Fig:Qruiser} indicates
that $\hat p_{A_{\rm{f.d.}}}^{\rm{(ren)}}$ performs best of all estimators.
Among the estimators
utilizing $d_i^{\rm{(un)}}+d_i^{\rm{(out)}}$, $\hat p_A^{\rm{(ren)}}$ has the smallest overall bias. Moreover, the variance of $\hat p_A^{\rm{(ren)}}$ is smaller than for
$\hat p_A^{\rm{(outdeg)}}$ for all properties, in particular the civil status.

\section{DISCUSSION AND CONCLUSIONS}\label{Sec:Discussion}

We developed statistical procedures for sampling vertices in social networks to account for the empirical fact that social networks generally include non-reciprocal
edges. The proposed estimation procedures typically outperformed existing methods that neglect directed edges. Among the scenarios investigated in the present study, the
best accuracy of estimation was obtained when undirected, in-directed, and out-directed degree are separately observed for sampled individuals. In the more realistic
scenario in which one only knows the sum of undirected and out-directed edges of sampled individuals, all estimation procedures are less precise. Our simulations also
showed that estimators of population proportions were highly sensitive to how
the property $A$ is allocated in the social network.

If the full directed degree $(d_i^{\rm{(un)}},d_i^{\rm{(in)}},d_i^{\rm{(out)}})$ is observed and the moments of the
degree distributions are known, our estimator $\hat\pi^{\rm{(ren)}}_{\rm{f.d.}}$ is compared to
$\hat\pi^{\rm{(indeg)}}$. It can be seen in Tables~\ref{Tab:ERObserveAll} and \ref{Tab:Qruiser}, and Figure~\ref{Fig:PL}
that $\hat\pi^{\rm{(ren)}}_{\rm{f.d.}}$ performs slightly better than $\hat\pi^{\rm{(indeg)}}$ in all the studied
situations. The corresponding estimated proportions given by $\hat p_{A_{\rm{f.d.}}}^{\rm{(ren)}}$ and $\hat
p_A^{\rm{(outdeg)}}$ in Figures~\ref{Fig:ERProportions}, \ref{Fig:PLProp}, and \ref{Fig:Qruiser} are very similar.

If only the out-degree $d_i^{\rm{(un)}}+d_i^{\rm{(out)}}$ is observed, we compare $\hat\pi^{\rm{(ren)}}$ and
$\hat\pi^{\rm{(outdeg)}}$ (Tables~\ref{Tab:ERObserveOutdegreeEstimatedParameters} and \ref{Tab:Qruiser}, and
Figure~\ref{Fig:PL}). We also include $\hat\pi^{\rm{(ren)}}_{\alpha,\lambda}$ in the comparison on the generated
networks, and it can be seen that the performance of $\hat\pi^{\rm{(ren)}}_{\alpha,\lambda}$ is only slightly better
than that of $\hat\pi^{\rm{(ren)}}$. Our estimator $\hat\pi^{\rm{(ren)}}$ outperforms $\hat\pi^{\rm{(outdeg)}}$ except
when the fraction of directed edges $\alpha$ is small (0.1 in Table~\ref{Tab:ERObserveOutdegreeEstimatedParameters}
and 0.25 in Figure~\ref{Fig:PL}). This corresponds to that $\hat\pi^{\rm{(ren)}}$ will deviate further from
$\hat\pi^{\rm{(outdeg)}}$ as $\alpha$ increases (Eq.~\eqref{Eq:Pi-kEstDegEstPar}). Figures~\ref{Fig:ERProportions}
and \ref{Fig:PLProp} indicate that the results of the estimators $\hat p_A^{\rm{(ren)}}$, $\hat
p_{A_{\alpha,\lambda}}^{\rm{(ren)}}$, and $\hat p_A^{\rm{(outdeg)}}$ depend much on the allocation of the
property $A$. We believe that it is of interest to further study how properties are distributed in empirical social
networks. 

If $\alpha$ is known, we can estimate $\lambda$ using only the mean sample out-degree in
Eq.~\eqref{Eq:LambdaHat}. Although generally difficult, it is possible to assess the fraction of directed edges in the
social network of a hidden population through direct methods. In many RDS studies, participants are asked questions that
experimenters use to quantify the nature of the relationship between a participant and its recruiter, e.g., friends, acquantiances or strangers
\citep[e.g.,][]{Ramirez2005,Wang2007,Ma2007}. With these questions, the authors aim to control for non-reciprocated relationships, which could lead to the participant
being excluded from the sample. This type of questions is also useful for assessing the directedness of the social network, because the fraction of coupons given by
strangers could be a measure of (non-)reciprocity. In \citet{Gile2012diagnostics}, another type of question more directly assessing reciprocation is suggested, e.g. ``Do
you think that the person to whom you gave a coupon would have given you a coupon if you had not participated in the study first?''. Another possible method to
estimate $\alpha$ would be to obtain information on the number of revisits $m$ used in Eq.~\eqref{Eq:AlphaHat}. This could be done by asking for
example ``Would you give a coupon to the person who gave you a coupon if he or she had not yet participated in the study?''.

The main focus of the present paper was on accounting for directed edges in a social network. There are also other assumptions in existing estimation procedures
(including the current one) worthy of relaxing. For example, the methods typically assume that participants choose coupon recipents uniformly at random among their
neighbors in the social network. In reality, they probably sample closely
connected neighbors more likely, which may bias estimators of selection
probabilities.
Extending the RDS methods by allowing weighted edges warrants for future work. It should be noted that our methods allow the two weights on the same
undirected edge in the opposite directions to be different, because our framework targets directed networks.

Random walks on directed networks have numerous other applications, including identification of important vertices
\citep{Brin1998conf,Langville2006book,Noh2004PRL,Newman2005SocNetw} and community detection \citep{Rosvall2008PNAS}. Therefore, we also hope that this work may contribute
to an increased understanding in other areas of network research that use random walks on directed networks.

\bibliographystyle{apalike}
\bibliography{rwdn}

\begin{thebibliography}{}

\bibitem[Boldi et~al., 2011]{Boldi2011}
Boldi, P., Rosa, M., Santini, M., and Vigna, S. (2011).
\newblock Layered label propagation: A multiresolution coordinate-free ordering
  for compressing social networks.
\newblock In {\em Proceedings of the 20th international conference on World
  Wide Web}, pages 587--596. ACM.

\bibitem[Boldi and Vigna, 2004]{Boldi2004}
Boldi, P. and Vigna, S. (2004).
\newblock The webgraph framework i: compression techniques.
\newblock In {\em Proceedings of the 13th international conference on World
  Wide Web}, pages 595--602. ACM.

\bibitem[Brin and Page, 1998]{Brin1998conf}
Brin, S. and Page, L. (1998).
\newblock Anatomy of a large-scale hypertextual web search engine.
\newblock {\em Proceedings of the Seventh International World Wide Web
  Conference}, pages 107--117.

\bibitem[Chung and Lu, 2002]{Chung2002PNAS}
Chung, F. and Lu, L.~Y. (2002).
\newblock {The average distances in random graphs with given expected degrees}.
\newblock {\em Proc. Natl. Acad. Sci. USA}, 99:15879--15882.

\bibitem[Chung et~al., 2003]{Chung2003PNAS}
Chung, F., Lu, L.~Y., and Vu, V. (2003).
\newblock {Spectra of random graphs with given expected degrees}.
\newblock {\em Proc. Natl. Acad. Sci. USA}, 100:6313--6318.

\bibitem[Donato et~al., 2004]{Donato2004EPJB}
Donato, D., Laura, L., Leonardi, S., and Millozzi, S. (2004).
\newblock {Large scale properties of the Webgraph}.
\newblock {\em Eur. Phys. J. B}, 38:239--243.

\bibitem[Doyle and Snell, 1984]{Doyle1984book}
Doyle, P.~G. and Snell, J.~L. (1984).
\newblock {\em Random Walks and Electric Networks}.
\newblock Math. Asso. Amer.

\bibitem[Erd\H{o}s and Renyi, 1960]{Erdos}
Erd\H{o}s, P. and Renyi, A. (1960).
\newblock On the evolution of random graphs.
\newblock {\em Publ. Math. Inst. Hungar. Acad. Sci}, 5:17--61.

\bibitem[Fortunato et~al., 2008]{FortunatoEtal}
Fortunato, S., Bogu{\~n}{\'a}, M., Flammini, A., and Menczer, F. (2008).
\newblock Approximating pagerank from in-degree.
\newblock In {\em Algorithms and Models for the Web-Graph}, pages 59--71.
  Springer.

\bibitem[Ghoshal and Barab\'{a}si, 2011]{Ghoshal2011NatComm}
Ghoshal, G. and Barab\'{a}si, A.~L. (2011).
\newblock Ranking stability and super-stable nodes in complex networks.
\newblock {\em Nat. Comm.}, 2:394.

\bibitem[Gile, 2011]{Gile2011JASA}
Gile, K.~J. (2011).
\newblock Improved inference for respondent-driven sampling data with
  application to hiv prevalence estimation.
\newblock {\em Journal of the American Statistical Association}, 106(493).

\bibitem[Gile and Handcock, 2010]{Gile2010SocMet}
Gile, K.~J. and Handcock, M.~S. (2010).
\newblock Respondent-driven sampling: An assessment of current methodology.
\newblock {\em Sociological Methodology}, 40(1):285--327.

\bibitem[Gile and Handcock, 2011]{Gile2011arXiv}
Gile, K.~J. and Handcock, M.~S. (2011).
\newblock Network model-assisted inference from respondent-driven sampling
  data.
\newblock {\em arXiv preprint arXiv:1108.0298}.

\bibitem[Gile et~al., 2012]{Gile2012diagnostics}
Gile, K.~J., Johnston, L.~G., and Salganik, M.~J. (2012).
\newblock Diagnostics for respondent-driven sampling.
\newblock {\em arXiv preprint arXiv:1209.6254}.

\bibitem[Goel and Salganik, 2010]{Goel2010}
Goel, S. and Salganik, M.~J. (2010).
\newblock Assessing respondent-driven sampling.
\newblock {\em Proceedings of the National Academy of Sciences},
  107(15):6743--6747.

\bibitem[Goh et~al., 2001]{Goh2001PRL}
Goh, K.~I., Kahng, B., and Kim, D. (2001).
\newblock {Universal behavior of load distribution in scale-free networks}.
\newblock {\em Phys. Rev. Lett.}, 87:278701.

\bibitem[Gong et~al., 2013]{GongEtal2013}
Gong, N.~Z., Xu, W., and Song, D. (2013).
\newblock Reciprocity in social networks: Measurements, predictions, and
  implications.
\newblock {\em arXiv preprint arXiv:1302.6309}.

\bibitem[Heckathorn, 1997]{Heckathorn1997}
Heckathorn, D.~D. (1997).
\newblock Respondent-driven sampling: a new approach to the study of hidden
  populations.
\newblock {\em Social problems}, pages 174--199.

\bibitem[Killworth and Bernard, 1976]{Killworth1976}
Killworth, P.~D. and Bernard, H.~R. (1976).
\newblock Informant accuracy in social network data.
\newblock {\em Human Organization}, 35(3):269--286.

\bibitem[Kwak et~al., 2010]{Kwak}
Kwak, H., Lee, C., Park, H., and Moon, S. (2010).
\newblock What is twitter, a social network or a news media?
\newblock In {\em Proceedings of the 19th international conference on World
  wide web}, pages 591--600. ACM.

\bibitem[Langville and Meyer, 2006]{Langville2006book}
Langville, A.~N. and Meyer, C.~D. (2006).
\newblock {\em Google's PageRank and beyond}.
\newblock Princeton University Press, Princeton.

\bibitem[Levin et~al., 2009]{Levin}
Levin, D.~A., Peres, Y., and Wilmer, E.~L. (2009).
\newblock {\em Markov chains and mixing times}.
\newblock Amer Mathematical Society.

\bibitem[Lov\'{a}sz, 1993]{Lovasz1993Boyal}
Lov\'{a}sz, L. (1993).
\newblock Random walks on graphs: A survey.
\newblock {\em Boyal Society Math. Studies}, 2:1--46.

\bibitem[Lu et~al., 2012]{Lu2012JRSS}
Lu, X., Bengtsson, L., Britton, T., Camitz, M., Kim, B.~J., Thorson, A., and
  Liljeros, F. (2012).
\newblock The sensitivity of respondent-driven sampling.
\newblock {\em Journal of the Royal Statistical Society: Series A (Statistics
  in Society)}, 175(1):191--216.

\bibitem[Lu et~al., 2013]{LuEtal}
Lu, X., Malmros, J., Liljeros, F., and Britton, T. (2013).
\newblock Respondent-driven sampling on directed networks.
\newblock {\em Electronic Journal of Statistics}, 7:292--322.

\bibitem[Ma et~al., 2007]{Ma2007}
Ma, X., Zhang, Q., He, X., Sun, W., Yue, H., Chen, S., Raymond, H.~F., Li, Y.,
  Xu, M., Du, H., et~al. (2007).
\newblock Trends in prevalence of hiv, syphilis, hepatitis c, hepatitis b, and
  sexual risk behavior among men who have sex with men: results of 3
  consecutive respondent-driven sampling surveys in beijing, 2004 through 2006.
\newblock {\em JAIDS Journal of Acquired Immune Deficiency Syndromes},
  45(5):581--587.

\bibitem[Magnus et~al., 2013]{Magnus2013}
Magnus, M., Kuo, I., Phillips~II, G., Rawls, A., Peterson, J., Montanez, L.,
  West-Ojo, T., Jia, Y., Opoku, J., Kamanu-Elias, N., et~al. (2013).
\newblock Differing hiv risks and prevention needs among men and women
  injection drug users (idu) in the district of columbia.
\newblock {\em Journal of Urban Health}, pages 1--10.

\bibitem[Masuda and Ohtsuki, 2009]{MasudaOhtsuki2009NewJPhys}
Masuda, N. and Ohtsuki, H. (2009).
\newblock Evolutionary dynamics and fixation probabilities in directed
  networks.
\newblock {\em New J. Phys.}, 11:033012.

\bibitem[Mislove et~al., 2007]{Mislove}
Mislove, A., Marcon, M., Gummadi, K.~P., Druschel, P., and Bhattacharjee, B.
  (2007).
\newblock Measurement and analysis of online social networks.
\newblock In {\em Proceedings of the 7th ACM SIGCOMM conference on Internet
  measurement}, pages 29--42. ACM.

\bibitem[Montealegre et~al., 2013]{Montealegre2013}
Montealegre, J.~R., Risser, J.~M., Selwyn, B.~J., McCurdy, S.~A., and Sabin, K.
  (2013).
\newblock Effectiveness of respondent driven sampling to recruit undocumented
  central american immigrant women in houston, texas for an hiv behavioral
  survey.
\newblock {\em AIDS and Behavior}, 17(2):719--727.

\bibitem[Moreno et~al., 1960]{Moreno}
Moreno, J.~L. et~al. (1960).
\newblock {\em The Sociometry Reader}.
\newblock Free Press New York.

\bibitem[Newman, 2010]{Newman2010}
Newman, M. (2010).
\newblock {\em Networks: an introduction}.
\newblock OUP Oxford.

\bibitem[Newman et~al., 2002]{Newman2002}
Newman, M.~E., Forrest, S., and Balthrop, J. (2002).
\newblock Email networks and the spread of computer viruses.
\newblock {\em Physical Review E}, 66(3):035101.

\bibitem[Newman, 2005]{Newman2005SocNetw}
Newman, M. E.~J. (2005).
\newblock {A measure of betweenness centrality based on random walks}.
\newblock {\em Soc. Netw.}, 27:39--54.

\bibitem[Noh and Rieger, 2004]{Noh2004PRL}
Noh, J.~D. and Rieger, H. (2004).
\newblock Random walks on complex networks.
\newblock {\em Phys. Rev. Lett.}, 92:118701.

\bibitem[Ramirez-Valles et~al., 2005]{Ramirez2005}
Ramirez-Valles, J., Heckathorn, D.~D., V{\'a}zquez, R., Diaz, R.~M., and
  Campbell, R.~T. (2005).
\newblock From networks to populations: the development and application of
  respondent-driven sampling among idus and latino gay men.
\newblock {\em AIDS and Behavior}, 9(4):387--402.

\bibitem[Resnick, 1992]{Resnick}
Resnick, S.~I. (1992).
\newblock {\em Adventures in Stochastic Processes}.
\newblock Birkhauser.

\bibitem[Rosvall and Bergstrom, 2008]{Rosvall2008PNAS}
Rosvall, M. and Bergstrom, C.~T. (2008).
\newblock {Maps of random walks on complex networks reveal community
  structure}.
\newblock {\em Proc. Natl. Acad. Sci. USA}, 105:1118--1123.

\bibitem[Rybski et~al., 2009]{Rybski2009}
Rybski, D., Buldyrev, S.~V., Havlin, S., Liljeros, F., and Makse, H.~A. (2009).
\newblock Scaling laws of human interaction activity.
\newblock {\em Proceedings of the National Academy of Sciences},
  106(31):12640--12645.

\bibitem[Salganik and Heckathorn, 2004]{Salganik2004}
Salganik, M.~J. and Heckathorn, D.~D. (2004).
\newblock Sampling and estimation in hidden populations using respondent-driven
  sampling.
\newblock {\em Sociological methodology}, 34(1):193--240.

\bibitem[Tang et~al., 2013]{Tang2013}
Tang, W., Huan, X., Mahapatra, T., Tang, S., Li, J., Yan, H., Fu, G., Yang, H.,
  Zhao, J., and Detels, R. (2013).
\newblock Factors associated with unprotected anal intercourse among men who
  have sex with men: Results from a respondent driven sampling survey in
  nanjing, china, 2008.
\newblock {\em AIDS and behavior}, pages 1--8.

\bibitem[Thompson, 2012]{ThompsonSampling}
Thompson, S.~K. (2012).
\newblock {\em Sampling}.
\newblock Wiley.

\bibitem[Tomas and Gile, 2011]{Tomas2011EJS}
Tomas, A. and Gile, K.~J. (2011).
\newblock The effect of differential recruitment, non-response and
  non-recruitment on estimators for respondent-driven sampling.
\newblock {\em Electronic Journal of Statistics}, 5:899--934.

\bibitem[Volz and Heckathorn, 2008]{Volz2008}
Volz, E. and Heckathorn, D.~D. (2008).
\newblock Probability based estimation theory for respondent driven sampling.
\newblock {\em Journal of Official Statistics}, 24(1):79.

\bibitem[Wang et~al., 2007]{Wang2007}
Wang, J., Falck, R.~S., Li, L., Rahman, A., and Carlson, R.~G. (2007).
\newblock Respondent-driven sampling in the recruitment of illicit stimulant
  drug users in a rural setting: Findings and technical issues.
\newblock {\em Addictive behaviors}, 32(5):924--937.

\bibitem[Wasserman and Faust, 1994]{Wasserman1994book}
Wasserman, S. and Faust, K. (1994).
\newblock {\em Social Network Analysis}.
\newblock Cambridge University Press, New York.

\bibitem[Wejnert, 2009]{Wejnert2009}
Wejnert, C. (2009).
\newblock An empirical test of respondent-driven sampling: Point estimates,
  variance, degree measures, and out-of-equilibrium data.
\newblock {\em Sociological methodology}, 39(1):73--116.

\end{thebibliography}

\end{document}